\documentclass[journal]{IEEEtran}

\newcommand{\UC}{\textup{}}

\newcommand{\UCM}{\textup{M}}

\newcommand{\UCMS}{\textup{MS}}

\newcommand{\UUC}{\text{Ucomp}}

\newcommand{\UUCM}{\text{UcompM}}
\newcommand{\UUCMI}{\text{UcompM1}}
\newcommand{\UUCMc}{\text{UcompMc}}
\newcommand{\UUCMS}{\text{UcompMS}}

\usepackage{cite}
\usepackage{graphicx}
\usepackage{epsfig,psfrag,subfigure}
\usepackage{array,multirow}
\usepackage{amsmath}
\usepackage{amssymb}
\usepackage{tikz}
\usepackage{algorithm,algpseudocode}
\usepackage{amsthm}

\newtheorem{theorem}{Theorem}

\newtheorem{proposition}[theorem]{Proposition}
\newtheorem{corollary}[theorem]{Corollary}

\newcommand{\mc}{\mathcal}
\newcommand{\mb}{\mathbf}

\newcommand{\KK}{K}
\begin{document}
\IEEEoverridecommandlockouts
\title{Universal Compression of a Mixture of\\Parametric Sources with Side Information}

\author{Ahmad~Beirami,~\IEEEmembership{Member,~IEEE,}
        Liling~Huang,
                Mohsen~Sardari,
        and~Faramarz~Fekri,~\IEEEmembership{Senior~Member,~IEEE}
\vspace{-.15in}

%\thanks{%Manuscript received April 19, 2005; revised April 22, 2005.
%The material in this paper was presented in part at the 2012 IEEE International Symposium on Information Theory (ISIT 2012), July 2012, and the 51st Annual Allerton Conference on Communication, Control and Computing (Allerton 2013), Monticello, IL, Oct 2013.}

\thanks{A. Beirami was  with the School of Electrical and Computer Engineering, Georgia Institute of Technology. He is currently with the Department of Electrical and Computer Engineering, Duke University, Durham, NC 27708, USA. e-mail: (ahmad.beirami@duke.edu).}

\thanks{L. Huang is with the School of Electronic, Information and Electrical Engineering, Shanghai Jiao Tong University, Shanghai, China e-mail: sunny\textunderscore hll@sjtu.edu.cn.
}

\thanks{M. Sardari and F. Fekri are with the School
of Electrical and Computer Engineering, Georgia Institute of Technology, Atlanta,
GA, 30332, USA. e-mail: (\{mohsen.sardai,~fekri\}@ece.gatech.edu).}

\thanks{This material is based upon work supported by the National Science Foundation under Grant No. CNS-1017234.}

\thanks{This paper was presented in part at the  2013 IEEE International Conference on Computer Communications (INFOCOM 2013)~\cite{INFOCOM13}, and the 51st Annual Allerton Conference on Communication, Control, and Computing (Allerton 2013)~\cite{Allerton13}, and the 15th IEEE International Workshop on Signal Processing Advances in Wireless Communications (SPAWC 2014)~\cite{spawc14}.}
}

\maketitle
%\thispagestyle{empty}
%\pagestyle{empty}

%\vspace{-0.05in}
\begin{abstract}
This paper investigates the benefits of the side information on the universal compression of sequences from a mixture of $K$ parametric sources.
The output sequence of the mixture source is chosen from the source $i \in \{1,\ldots ,K\}$ with a $d_i$-dimensional parameter vector at random according to probability vector $\mb{w} = (w_1,\ldots,w_K)$. The average minimax redundancy of the universal compression of a new random sequence of length $n$ is derived when the encoder and the decoder have a common side information of $T$ sequences generated independently by the mixture source.  Necessary and sufficient conditions on the distribution $\mb{w}$ and the mixture parameter dimensions $\mb{d} = (d_1,\ldots,d_K)$ are determined such that the side information provided by the previous sequences results in a reduction in the first-order term of the average codeword length compared with the universal compression without side information. %The reduction is characterized using closed form expressions.
Further, it is proved that the optimal compression with side information corresponds to the clustering of the side information sequences  from the mixture source.
Then, a  clustering technique is presented to better utilize the side information by classifying the data sequences from a mixture source. Finally, the performance of the clustering on the universal compression with side information is validated using computer simulations on real network data traces.
\end{abstract}


\section{Introduction}
\label{sec:introduction}

\IEEEPARstart{U}{niversal} compression aims at reducing the average number of bits required to describe a sequence from an unknown source from a family of sources, while good performance is desired for most of the sources in the family~\cite{Davisson_noiseless_coding, LZ77, CTW95, Rissanen_universal_modeling, Feder_Hierarchical, Effros_BWT, Baron_O_n,Barron_Cover_91,KT_estimator}. However, it often needs to observe a very long sequence so that it can effectively learn the existing patterns in the sequence for efficient compression. Therefore, universal compression  performs poorly on relatively small sequences~\cite{ISIT11,Merhav_Feder_IT} where sufficient data is not available for learning of the statistics and training of the encoder.
On the other hand, the presence of side information at the decoder has proven to be useful in several source coding applications (cf.~\cite{slepian_wolf, wyner_ziv,CEO_problem} and the references therein). In particular, the impact of side information on {\em universal} compression has also been shown to be useful (cf.~\cite{Gallager-source-coding,KT_estimator,ISIT12_gain, ISIT12_distributed}).
However, to the best of the authors' knowledge, the problem of  the universal compression of a mixture of parametric sources with side information has not been explored in the literature.

The recent rapid growth in the network traffic has motivated new research directions to leverage the existing correlations in the sequences (network packets) in order to reduce the traffic. These solutions must be transparent to the user and the application and hence must reside on the network layer, where the correlated sequences in the network flow are present~\cite{SIVAREP,Siva11,INFOCOM12, TON14}.
As network packets are relatively small, universal compression solutions (if employed naively) do not result in much traffic reduction~\cite{ISIT11, INFOCOM12,INFOCOM13, TON14}.
Further, the existing universal compression schemes do not exploit the cross correlation among the packets destined to different users.
As such, recently, we proposed universal compression of network packets using network memory in~\cite{INFOCOM12, TON14}, where the common memory between the encoder server (or router) and the decoder router was used as the side information to improve the performance of universal compression on network packets. As each packet may be generated by a different source, a realistic modeling of the network traffic requires to consider the content server to be a mixture source~\cite{INFOCOM13}. This motivates us to study the universal compression of sequences from a mixture source using common side information between the encoder and the decoder.  With a different motivation, Krishnan and Baron recently proposed a MDL-based parallel universal compression algorithm to exploit the cross correlation among the packets~\cite{ISIT14_Baron, Baron_MDL}.

In~\cite{ITW11, ISIT12_distributed}, we derived the optimal universal compression performance with side information for a single source, i.e., $\KK=1$; we proved that significant improvement is obtained from the side information in the universal compression of small sequences when sufficiently large side information is available. It was shown that a few megabytes of side information can drive the sequence length very close to the entropy. 
On the other hand, it is natural to expect that network packets that can be observed on a router are generated by a mixture of parametric sources.
Motivated by this fact, in this paper, we extend the setup of the memory-assisted universal compression to a mixture of $\KK$ parametric sources. 
Although the problem formulation is inspired from the network traffic compression,
universal compression of a mixture source with side information
 finds applications in a wide variety of problems, such as data storage systems, and migration of virtual machines, where the compression of data before transmission is desirable.

\begin{figure}
\centering
%\vspace{0.1in}
\includegraphics[width = 0.35\textwidth]{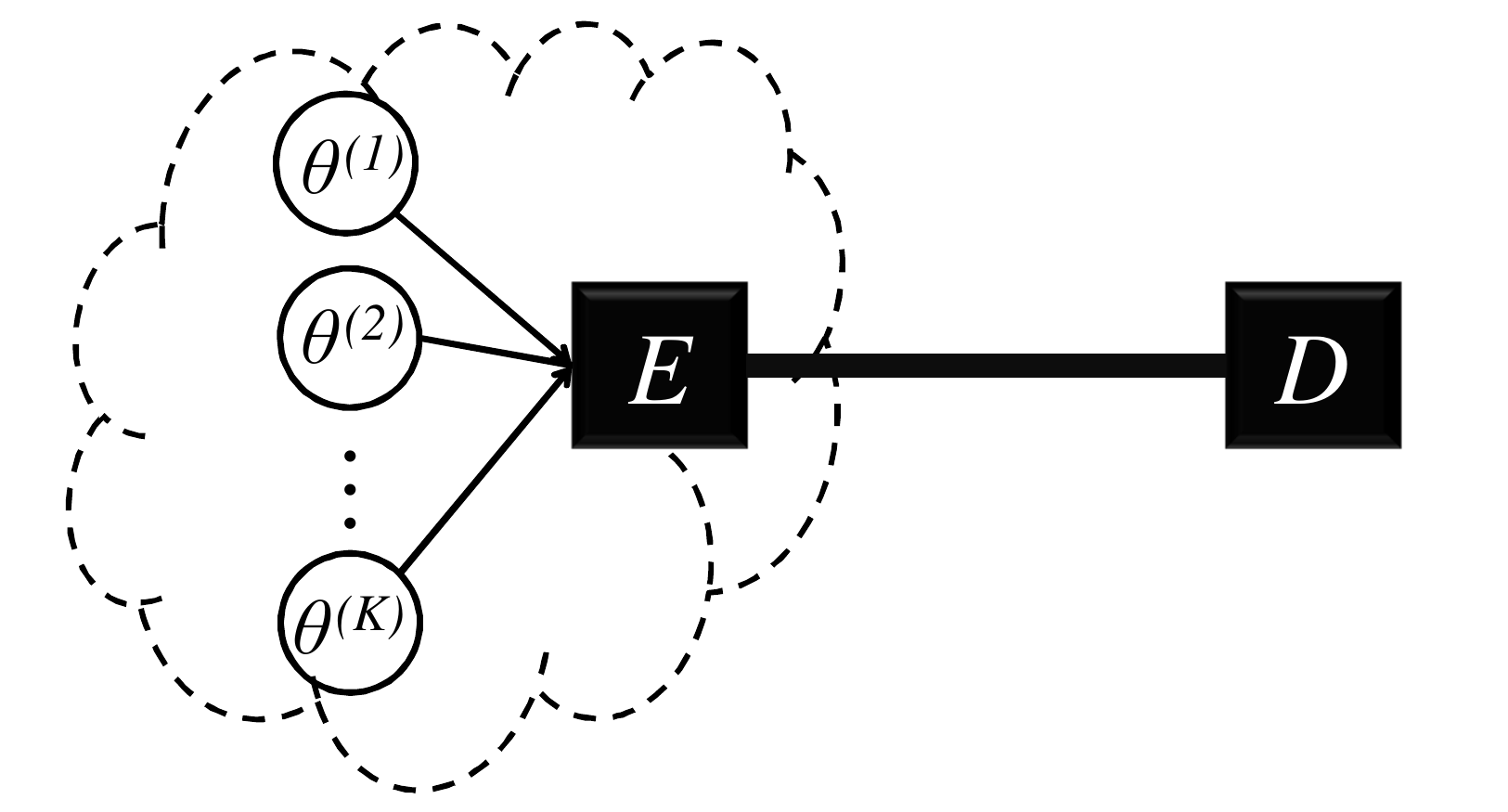}
\vspace{-0.05in}
\caption{The basic scenario of universal compression with side information for a mixture source.}
\vspace{-.2in}
\label{fig:single_hop}
\end{figure}

As shown in Fig.~\ref{fig:single_hop}, we assume that each sequence (e.g., network packet) is a sample of length $n$ from a mixture of $\KK$ parametric sources. We consider the scenario where $T$ sequences from the mixture source are shared as side information between the encoder $E$ and the decoder $D$ and the first objective is to derive the average minimax redundancy incurred in the {\em optimal} universal compression with side information as a function of $n$, $K$, and $T$.
We further develop a clustering algorithm for the universal compression with side information based on the Hellinger distance of the sequences and show its effectiveness on real network traffic traces.
We prove that the adopted clustering algorithm is consistent and asymptotically {\em optimal} using the side information in the sense that, given the side information, the minimum codeword length in the universal compression of a new sequence from the mixture source using side information is attained.

Our contributions in this paper can be summarized as follows:
\begin{itemize}
\item We formally characterize the average minimax redundancy incurred in universal compression of a random sequence of length $n$ from a mixture source given that the encoder and the decoder have access to a shared side information of $T$ sequences (each of length $n$ from the mixture of $K$ parametric sources).

\item We demonstrate that the performance of the optimal universal compression with side information (in the minimax sense) is almost surely that of the universal compression with perfect clustering of the memory based on the originator source in the mixture when source labels are available. Hence, clustering is optimal in the first-order redundancy term for universal compression with side information.

\item We propose a clustering strategy for the side information that aims at grouping the side information sequences that share similar statistical
properties. A newly generated packet by the mixture source is classified into one of the clusters for compression. We demonstrate the effectiveness of the proposed algorithm through experiments performed on real network traffic traces.
\end{itemize}

The rest of this paper is organized as follows. In Section~\ref{sec:background}, we review the necessary background on universal compression. In Section~\ref{sec:problem_setup},  we present the formal definition of the problem.
In Section~\ref{sec:entropy}, we derive the entropy of the mixture source, which serves as a lower limit on the average codeword length.
In Section~\ref{sec:universal}, we provide the main results on the universal compression of mixture sources with and without side information and discuss their implications. In Section~\ref{sec:cluster}, we present the clustering algorithms used for the compression of the mixture sources. In Section~\ref{sec:simulation}, we provide simulation results that support our theoretical results on the compression of the mixture sources. Finally, Section~\ref{sec:conclusion} concludes this paper.

\section{Background on Universal Source Coding}
\label{sec:background}
In this section, we briefly review the necessary background on the universal compression of parametric sources. We defer the generalization to a mixture source to Section~\ref{sec:problem_setup}.
 Let a parametric source be defined using a $d$-dimensional parameter vector $\theta = (\theta_1,...,\theta_d)\in \Lambda_d$ that is a priori unknown, where $d$ denotes the number of the source parameters and $\Lambda_d \subset  \mathbb{R}^d$ is the space of $d$-dimensional parameter vectors of interest.
Denote $\mu_\theta$ as the parametric source (i.e., the probability measure defined by the parameter vector $\theta$ on sequences of length $n$).

Let $\mc{X}$ denote a finite alphabet. Let $X^n$ denote a sample (random vector of length $n$) from the probability measure $\mu_\theta$. We further denote $x^n = (x_1,...,x_n) \in \mc{X}^n$ as a realization of the random vector $X^n$. Then, define $H_n(\theta) \triangleq H(X^n|\theta)$ as the source entropy given the parameter vector $\theta$, i.e.,
\begin{equation}
H_n(\theta) \hspace{-.02in}= \hspace{-.02in} \mb{E} \log \hspace{-.03in}\left(\frac{1}{\mu_\theta(X^n)} \right)\hspace{-.04in} =\hspace{-.02in}\sum_{x^n\in \mc{X}^n}\mu_{\theta}(x^n) \log \hspace{-.03in}\left(\frac{1}{\mu_\theta(x^n)}\right)\hspace{-.03in}.
\label{eq:entropy}
\end{equation}
Throughout this paper $\log(\cdot)$ always denotes the logarithm in base $2$ and expectations are taken over the random sequence $X^n$ with respect to the probability measure $\mu_\theta$.

In this paper, we focus on the class of strictly lossless uniquely decodable fixed-to-variable codes defined as the following.
The code $c_n: \mc{X}^n \to \{0,1\}^*$ is called strictly lossless (also called zero-error) on sequences of length $n$ if there exists a reverse mapping $d_n:\{0,1\}^* \to \mc{X}^n$ such that
$\forall x^n \in \mc{X}^n$, we have $d_n(c_n(x^n)) = x^n.$
Further, let $l:\mc{X}^n\to \mathbb{R}$ denote the universal strictly lossless length function for the codeword $c_n(x^n)$ associated with the sequence $x^n$ such that $l(\cdot)$ satisfies Kraft's inequality to ensure unique decodability. That is
$
\sum_{x^n \in \mc{X}^n}2^{-l(x^n)} \leq 1.
$
In this paper, we ignore the integer constraint on the length function, which results in a negligible redundancy upper bounded by $1$ bit analyzed exactly in~\cite{Precise_Minimax_Redundancy, huffman_redundancy}.

Denote $R_{n,d}(l,\theta)$  as the average (expected) redundancy of the code $c_n$ with length function $l$ on a sequence of length $n$ for the parameter vector $\theta$, defined as
\begin{equation}
R_{n,d}(l,\theta) = \mb{E}l(X^n) -  H_n(\theta).
\label{eq:redundancy-def}
\end{equation}
Note that the average redundancy is non-negative. Further, a (universal) code is called weakly optimal if its average codeword length normalized to the sequence length uniformly converges to the source entropy rate, i.e., $\lim_{n \to \infty} \frac{1}{n} R_{n,d}(l,\theta)= 0$ for all~$\theta\in \Lambda_d$.

Define $\underline{R}_{n,d}$ as the average maximin redundancy, i.e.,
\begin{equation}
\underline{R}_{n,d} = \max_{p(\cdot)} \min_{l} \int_{ \Lambda_d} R_{n,d}(l,\theta) p(\theta)d\theta.
\label{eq:def-maximin}
\end{equation}
The average maximin redundancy is associated with the best code under the worst prior on the space of parameter vectors (i.e., the capacity achieving Jeffreys' prior).
Let $\bar{R}_{n,d}$ denote the average minimax redundancy, which is defined as
\begin{equation}
\bar{R}_{n,d} = \min_{l} \max_{\theta}   R_{n,d}(l,\theta).
\label{eq:def-minimax}
\end{equation}
Gallager showed that the average minimax redundancy and the average maximin redundancy (as defined above) are equal~\cite{Gallager-source-coding}.
Let $\mc{I}(\theta)$ be the Fisher information matrix associated with the parameter vector $\theta$, i.e.,
\begin{equation}
\mc{I}(\theta)\hspace{-.03in} \triangleq \lim_{n \to \infty} \hspace{-.01in}\frac{1}{n\log e}\mb{E}\hspace{-.03in}\left\{\frac{\partial^2}{\partial \theta_i \partial \theta_j} \log \hspace{-.03in}\left( \frac{1}{\mu_\theta(X^n)}\right)\hspace{-.03in} \right\}\hspace{-.02in}.
\label{eq:fisher}
\end{equation}
Fisher information matrix quantifies the amount of information, on average, that each symbol in a sample sequence $x^n$ from the source conveys about the source parameter vector.
Let Jeffreys' prior on the parameter vector $\theta$ be denoted by
\begin{equation}
p_J(\theta) \triangleq \frac{|\mc{I}(\theta)|^{\frac{1}{2}}}{\int_{\Lambda_d}|\mc{I}(\lambda)|^\frac{1}{2}d\lambda}.
\label{eq:Jeffreys'}
\end{equation}
Jeffreys' prior is optimal in the sense that the average minimax redundancy is asymptotically achieved (up to a constant) when the parameter vector $\theta$ is assumed to follow Jeffreys' prior~\cite{Clarke_Barron,atteson_markov,Gallager-source-coding}.\footnote{The boundary risk is asymptotically strictly larger than the interior risk by a constant using Jeffreys' prior and when the space of the parameter vectors includes the boundary, a modification of Jeffreys' prior towards the boundary to compensate for this is minimax optimal (cf.~\cite{Minimax_Redundancy_Memoryless}).} Jeffreys' prior is particularly interesting because it is also maximin optimal, which corresponds to the worst-case prior for the best compression scheme (called the capacity achieving prior)~\cite{Gallager-source-coding}.

We need some regularity conditions to hold for the parametric model so that our results can be derived.
\begin{enumerate}%[label = P\arabic*.]
\item The parametric model is smooth, i.e., twice differentiable with respect to $\theta$ in the interior of $\Lambda$ so that the Fisher information matrix can be defined. Further, the limit in~\eqref{eq:fisher} exists.

\item The determinant of fisher information matrix is finite for all $\theta$ in the interior of $\Lambda$ and the normalization constant in the denominator of~\eqref{eq:Jeffreys'} is finite.

\item The parametric model has a minimal $d$-dimensional representation, i.e., $\mc{I}(\theta)$ is full-rank. Hence, $\mc{I}^{-1}(\theta)$ exists.

\item We require that the central limit theorem holds for the maximum likelihood estimator $\hat{\theta}(x^n)$ of each $\theta$ in the interior of $\Lambda$ so that $(\hat{\theta}(X^n) - \theta)\sqrt{n}$ converges to a normal distribution with zero mean and covariance matrix $\mc{I}^{-1}(\theta)$.
\end{enumerate}

The average minimax (maximin) redundancy is well studied for a single parametric source given by the following theorem.
\begin{theorem}[{\bf \cite{Clarke_Barron, atteson_markov}}]
The average minimax (maximin) redundancy is given by
\begin{equation}
\bar{R}_{n,d} = \frac{d}{2}  \log\left( \frac{n}{2\pi e} \right) + \log \int_{ \Lambda_d}
|\mc{I}_n(\theta)|^{\frac{1}{2}}d\theta + o(1).\footnote{Throughout this work, we have used the following asymptotic notations:
\begin{itemize}
  \item $f(n) = o(g(n))$ iff $|f(n)| \leq |g(n)| \epsilon,~\forall \epsilon$,
  \item $f(n) = O(g(n))$ iff $|f(n)| \leq |g(n)| k,~\exists k$,
  \item $f(n) = \omega(g(n))$ iff $g(n) = o(f(n))$,
   \item $f(n) = \Omega(g(n))$ iff $g(n) = O(f(n))$, 
  \item $f(n) \sim g(n)$ iff $\lim_{n \to \infty} f(n)/g(n)=1$,
  \item $f(n) \lesssim g(n)$ iff $f(n) = o(g(n))$, and
  \item $f(n) \gtrsim g(n)$ iff $f(n) = \omega(g(n))$.
\end{itemize}}
\label{eq:minimax}
\end{equation}
\label{thm:maximin}
\end{theorem}
According to Theorem~\ref{thm:maximin}, the average maximin redundancy scales as $\frac{d}{2} \log n+ O(1)$. This redundancy may indeed be a significant overhead on top of the entropy for small sequences, as the second term in~\eqref{eq:minimax} could be relatively large for small $n$ as characterized in~\cite{ISIT11}.

\section{Problem Setup}
\label{sec:problem_setup}
In this section, we present the setup of the universal compression with common side information at the encoder and the decoder.
Let $\Delta \triangleq \left\{\theta^{(i)}\right\}_{i=1}^\KK$ denote the set of $\KK \triangleq |\Delta|$  parameter vectors of interest where $\theta^{(i)} \in \Lambda_{d_i}$ is a $d_i$-dimensional parameter vector. Note that we let $\KK$ deterministically scale with $n$.
Let $d_{\text{max}} \triangleq \max\{d_1,\ldots,d_K\}$ denote the maximum dimension of the parameter vectors, where we assume that $d_{\text{max}} = O(1)$, i.e., $d_{\text{max}}$ is finite.
We further assume that for any $d<d'$, we have $\Lambda_d \subset \Lambda_{d'}$, and hence, $\Delta$ consists of $K$ points on the space $\Lambda_{d_{\text{max}}}$. 
In this setup, as in Fig.~\ref{fig:single_hop}, the source is a mixture of $\KK$ parametric sources $\mu_{\theta^{(1)}},\ldots ,\mu_{\theta^{(\KK)}}$, where for all $i \in [K] \triangleq \{1,\ldots,K\}$, $\theta^{(i)}$  is  a $d_i$-dimensional unknown parameter vector.
For the generation of each sequence of length $n$, the generator source is selected according to the probability vector $\mb{w} = (w_1,\ldots,w_K)$ from the mixture, i.e., $\Delta$.
In other words, $p(\theta|\Delta) = \sum_{i=1}^{\KK} w_i\delta(\theta - \theta^{(i)})$, %where $\theta^{(i)}$ follows Jeffreys' prior on $\Lambda_{d_i}$ and
 where $w_i$ is the probability that the sequence is generated by source $\theta^{(i)}$ in the mixture. The random set $\Delta$ (which is unknown a priori) is generated once and is used thereafter for the generation of all sequences from the mixture source.
Let $S$ be a random variable that determines the source index from which of the sequence is generated. As such, $S$ follows the distribution $\mb{w}$ over $[\KK]$, i.e., $\mb{P}[S=i] = w_i$.
Then, by definition, we have $\theta = \theta^{(S)}$ given $\Delta$. Unlike $\Delta$ that is generated once, $S$ is chosen via $\mb{w}$ every time a new sequence is generated.
Let the mixture entropy $H(\mb{w})$ be defined as $H(\mb{w}) = - \sum_{i\in[K]} w_i \log w_i$.\footnote{We define entropy $H(\mb{r})$ for any vector $\mb{r}$ such that $\sum_{i}r_i =1$ in the same manner throughout the paper.}

%We consider the following scenario. 
We assume that, in Fig.~\ref{fig:single_hop}, both the encoder $E$ and the decoder $D$ have access to a common side information of $T$ previous sequences (indexed by $[T]$) from the mixture of $\KK$ parametric sources, where each of these sequences is independently generated according to the above procedure.
Let $m \triangleq nT$ denote the aggregate length of the previous $T$ sequences from the mixture source.\footnote{For simplicity of the discussion, we consider the lengths of all sequences to be equal to $n$. However, most of the results are readily extendible to the case where the sequences are not necessarily equal in length.} Further, denote $\mb{y}^{n,T} = \left\{y^{n}(t)\right\}_{t=1}^{T}$ as the set of the previous $T$ sequences shared between $E$ and $D$, where $y^n(t)$ is a sequence of length $n$ generated from the source $\theta^{S(t)}$ at time epoch $t$, where $S(t)$ follows $\mb{w}$ on $[\KK]$. In other words, $y^n(t) \sim \mu_{\theta^{(S(t))}}$. Further, denote $\mb{S}$ as the vector $\mb{S} = \left(S(1),...,S({T})\right)$, which contains the indices of the sources that generated the $T$ previous side information sequences.

Let $\l_\text{M}(x^n,\mb{y}^{n,T})$ denote a length function that utilizes the side information $\mb{y}^{n,T}$ in the compression of a new sequence $x^n$.
The objective is to analyze the average redundancy in the compression of a new sequence $x^n$ that is independently generated by the same mixture source with source index $Z$ (which also follows $\mb{w}$).
We investigate the fundamental limits of the universal compression with side information ($\mb{y}^{n,T}$) that is shared between the encoder and the decoder and compare with that of the universal compression without side information of the previous sequences. 
In this respect, it is straightforward to show that the minimax and maximin average redundancy are equivalent and are given by the capacity of the channel between the sequence $X^n$ and the parameter vectors $\Delta$ given side information sequence $Y^{n,T}$. Hence, $I(X^n; \Delta|\mb{Y}^{n,T})$ and $I(X^n;\Delta)$, for different values of the sequence length $n$, memory (side information) size $m=nT$, the weight of the mixture $\mb{w}$, and the dimensions of the parameter vectors $\mb{d}$, serve as two of the main fundamental limits of the universal compression in this setup.

\section{Entropy of the Mixture Source: Compression with Known Source Parameter Vectors}
\label{sec:entropy}
In this section, we derive the limits of compression when the source parameter vectors are known.
It is well known that for the mixture source, optimal compression is achieved by mixing  the models. In other words, let $p(x^n)$ denote the mixture probability distribution on sequences of length $n$, which is defined as
\begin{equation}
p(x^n) = \sum_{i=1}^K w_i \mu_{\theta^{(i)}}(x^n).
\end{equation}
Hence, the length function 
\begin{equation}
l(x^n) = \log \left( \frac{1}{p(x^n)}\right)
\end{equation}
is the optimal length function in this case, and it will achieve the entropy of the mixture source. 

To derive the limits of compression for known source parameter vectors, we need to derive the entropy of the mixture source. 
Let $H_n(\Delta,Z) \triangleq H(X^n|\Delta,Z)$ be defined as the entropy of a random sequence $X^n$ from the mixture source given that the source parameters are known to be the set $\Delta$ and the index of the source that has generated the sequence (i.e., $Z$) is also known.\footnote{We assume that the random set of parameter vectors is generated once and used for the generation of all sequences of length $n$ thereafter. Therefore, throughout the paper, whenever we assume that $\Delta$ is given, we mean that the set of the parameter vectors is known to be the set $\Delta$.}  %In other words, the parameter vector $\theta^{(Z)}$ associated with sequence $X^n$ is known. 
Then, in this case, by definition
\begin{equation}
H_n(\Delta,Z) = \sum_{i=1}^\KK w_i H_n(\theta^{(i)}),
\label{eq:base_entropy}
\end{equation}
where $H_n(\theta^{(i)})$ is the entropy of source $\mu_{\theta^{(i)}}$ given $\theta^{(i)}$ defined in~\eqref{eq:entropy}.
Note that $H_n(\Delta,Z)$ is {\em not} the achievable performance of the compression. It is merely introduced here so as to make the presentation of the results more convenient.

Let the set $\Delta$ be partitioned into subsets in the following fashion.
\begin{equation}
\Delta = \cup_{d=1}^{d_{\text{max}}} \Delta_d,
\end{equation}
where $\Delta_d$ is the set of the $d$-dimensional parameter vectors in $\Delta$. Further, let $K_d \triangleq |\Delta_d|$ be the number of parameter vectors in set $\Delta_d$.
In other words, $K_d$ is the number of sources of dimension $d$ in the mixture source.
Hence, $\sum_{d=1}^{d_{\text{max}}}
 K_d = K$. Now, we can relabel the elements in $\Delta$ according to their parameter vectors. Let $\Delta_d = \{\theta^{(d,1)},\ldots,\theta^{(d,K_d)} \}.$ Denote $\mb{w}_d = (w_{d,1},\ldots,w_{d,K_d})$ as the weight of the $d$-dimensional parameter vectors. Further, let $v_d\triangleq \sum_{i=1}^{K_d} w_{d,i}$ be the aggregate weight of all $d$-dimensional parameter vectors and denote $\mb{v} \triangleq (v_1,\ldots,v_{d_{\text{max}}})$. Let $\mb{\hat{w}}_{d} \triangleq \mb{w}_{d}/v_d$, i.e., we have $\hat{w}_{d,i} \triangleq w_{d,i}/v_d$, for $1\leq i\leq K_d$.

 Hence, $H_n(\Delta,Z)$ can be rewritten as
 \begin{align}
H_n(\Delta,Z) & = \sum_{d=1}^{d_{\text{max}}} \sum_{i=1}^{K_d} w_{d, i} H_n(\theta^{(d,i)})\\
& = \sum_{d=1}^{d_{\text{max}}} v_d \sum_{i=1}^{K_d} {\hat{w}}_{d, i} H_n(\theta^{(d,i)}).
 \end{align}

Next, we derive the entropy of the mixture source (which sets the asymptotic fundamental lower limit on the codeword length for the known source parameters case), i.e., when $\Delta$ is known. Define $H_n(\Delta) \triangleq H(X^n|\Delta)$.
\begin{theorem}
The entropy of the mixture source for all $\Delta$ except for a set $A(n)$ whose volume asymptotically vanishes as $n \to \infty$, is given by
\begin{equation}
H_n(\Delta) = H_n(\Delta,Z) + H(\mb{v}) + \sum_{d=1}^{d_\text{max}} v_d H_d+ o(1),
%~a.s.,\footnote{An event $A$ happens a.s. (almost surely) if and only if $\mathbb{P}[A] =1$.}
%^,\footnote{Please note that the sample space is the set of all source parameter vectors $\Delta = \left\{\theta^{(i)}\right\}_{i=1}^{\KK}$ such that $\theta^{(i)}$ is drawn independently from Jeffreys' prior.}
\end{equation}
where $H_d$ is given by
\begin{equation}
H_d = \left\{
\hspace{-.05in}
\begin{array}{ll}
  H(\mb{\hat{w}}_d) & \text{if~} H(\mb{\hat{w}}_d) \lesssim \frac{d}{2} \log n \\
  \bar{R}_{n,d} & \text{if~} H(\mb{\hat{w}}_d) \gtrsim \frac{d}{2} \log n
 \end{array}
\right. ,
\label{eq:Hd}
\end{equation}
and $\bar{R}_{n,d}$ is given by~\eqref{eq:minimax}.
\label{thm:entropy_rate}
%\vspace{-.1in}
\end{theorem}
\begin{IEEEproof}
The proof is explained in the appendix.
\end{IEEEproof}
Theorem~\ref{thm:entropy_rate} determines the entropy of the mixture source, which corresponds to the minimum codeword length when the parameter vectors in the set $\Delta$ are known to the encoder and the decoder (i.e., non-universal compression). Note that $H_n(\Delta)$ also serves as a trivial lower bound on the codeword length for the case of universal compression (i.e., unknown parameter vectors).
For sufficiently low-entropy $\hat{\mb{w}}_d$ (or roughly sufficiently small $K_d$), the price of describing the $d$-dimensional parameter vectors is, on average, equal to $H(\mb{\hat{w}}_d)$, which corresponds to describing the respective source parameter vector in the encoder.

\vspace{0.12in}
{\noindent \bf Remark.} Theorem~\ref{thm:entropy_rate} does not hold for an asymptotically vanishing volume of the parameter vectors. This is because one can choose the parameter vectors in a way that they do not conform to asymptotic scaling. For example, if all the parameter vectors are chosen to be equal, then the extra redundancy term over $H_n(\Delta,Z) = H_n(\theta^{(1)})$ would be zero. On the other hand, the result states that the volume of the space covered by such choices would become vanishingly small as $n \to \infty$. This is equivalent to saying if the parameter vectors are chosen independently according to a uniform prior on the state of parameter vectors, then the probability of the event that they do not conform to the scaling predicted by Theorem~\ref{thm:entropy_rate} is vanishingly small.

The following corollary describes the entropy when the number of source parameter vectors are sufficiently small.
\begin{corollary}
\label{thm:small_K}
If $K = O\left(n^{\frac{1}{2}-\epsilon}\right)$ for some $\epsilon > 0$, then for all $\Delta$ except for a set $A(n)$ whose volume asymptotically vanishes as $n \to \infty$, we have
\begin{equation}
H_n(\Delta) = H_n(\Delta,Z) + H(\mb{w})+ o(1).
\end{equation}
\end{corollary}
\begin{IEEEproof}
Since $K =  O\left(n^{\frac{1}{2}-\epsilon}\right)$ for some $\epsilon > 0$, we have  $K_d =  O\left(n^{\frac{d}{2}-\epsilon}\right)$ for some $\epsilon > 0$ and for all $1\leq d \leq d_\text{max}$. Thus, we have $H(\mb{\hat{w}}_d)  \lesssim \frac{d}{2} \log n$. Thus, $H_d =  H(\mb{\hat{w}}_d)$ for all $1\leq d\leq d_{\text{max}}$.
The proof is completed by noting that 
\begin{equation}
H(\mb{w}) = H(\mb{v}) +  \sum_{d=1}^{d_\text{max}} v_d H(\mb{\hat{w}}_d).
\end{equation}
\end{IEEEproof}
According to the corollary, when $K = O\left(n^{\frac{1}{2}-\epsilon}\right)$ for some $\epsilon > 0$, the optimal coding strategy (when the source parameters are known) for asymptotically almost all the parameter vectors would be to encode the source index $Z$ and then use the optimal code (e.g., Huffman code) associated with parameter $\theta^{(Z)}$ for sequences of length $n$ to encode the sequence $x^n$.
In fact, if $H(\mb{w}) \lesssim \frac{d}{2} \log n $, then the cost of encoding the parameter is asymptotically smaller than the cost of universally encoding the parameter and hence it is beneficial to encode the parameter vector using an average of $H(\mb{w})$ bits.
Further, if $\KK=1$, then $\Delta=\theta^{(1)}$ and $Z=1$ would be deterministic. Hence, $H_n(\Delta) = H_n(\Delta,Z) = H_n(\theta^{(1)})$, which was introduced in~\eqref{eq:entropy} as the average compression limit for the case of a single known source parameter vector.

\begin{corollary}
\label{thm:big_K}
If $H(\mb{\hat{w}}_d)  \gtrsim \frac{d}{2} \log n$ for all $1\leq d \leq d_\text{max}$  such that $v_d >0$, then for all $\Delta$ except for a set $A(n)$ whose volume asymptotically vanishes as $n \to \infty$, we have
\begin{equation}
H_n(\Delta) = H_n(\Delta,Z) + H(\mb{v})+  \sum_{d=1}^{d_\text{max}} v_d \bar{R}_{n,d}+o(1).
\end{equation}
\end{corollary}
\begin{IEEEproof}
The proof is very similar to the previous  corollary and is omitted for brevity.
\end{IEEEproof}
According to the corollary, in the case where the number of sources in the mixture is very large, the mixture entropy converges to $H_n(\Delta,Z)$ plus $H(\mb{v})$ plus the weighted average of the $\bar{R}_{n,d}$ terms (which are exactly the average maximin redundancy in the {\em universal} compression of parametric sources with $d$ {\em unknown} parameters given in Theorem~\ref{thm:maximin}).
At the first glance, it may seem odd that the codeword length in the case of {\em known} source parameter vectors incurs a term that is associated with the universal compression of a source with an {\em unknown} parameter vector. A closer look, however, reveals that in this case the cost of encoding the source index of a $d$-dimensional parameter vector surpasses the cost of universally encoding the source parameter vector. Hence, intuitively, it no longer makes sense to encode the $d$-dimensional parameter vector for the compression of the sequence $x^n$ using an average of $H(\mb{\hat{w}}_d)$ bits.
More rigorously speaking, as was shown in the proof of Theorem~\ref{thm:entropy_rate}, the probability distribution of $x^n$ given $\theta \in \Delta_d$ would converge to the probability distribution of $x^n$ when the source has one {\em unknown} $d$-dimensional parameter vector that follows Jeffreys' prior. This in turn results in the $\bar{R}_{n,d}$ term in the compression performance.

\section{Fundamental Limits of Universal Compression for Mixture Sources}
\label{sec:universal}
In the previous section, we derived the limits of the compression of mixture sources when the source parameter vectors are known. In this section, we will turn to the universal compression problem and will quantify the benefits of side information. 
%In this section, we state our main results on the fundamental limits of universal compression for mixture sources with and without side information.  %The proofs are deferred to Section~\ref{sec:analysis}.
To see the impact of the universality and side information on the compression performance, i.e., to investigate the impact of $\Delta$ being unknown, we will need to analyze and compare the average minimax redundancy (the excess codeword length on top of the entropy) for the following important {\em fundamental} schemes.
\begin{itemize}
\item {\UUC}: Universal compression, which is the conventional compression based solution. This is the usual universal compression in the literature with length function $l(x^n)$.
\item {\UUCM}: Universal compression with side information (common memory between the encoder and the decoder), which takes in the side information sequence into the encoding and decoding with length function $l_\text{M}(x^n,\mb{y}^{n,T})$.
\item {\UUCMS}: Universal compression with side information and source indices, which uses the side information sequences and also the indices of the sources that generated them (at the encoder/decoder). The respective length function will be denoted by $l_\text{MS}(x^n, \mb{y}^{n,T}, \mb{S}, Z)$.\footnote{{\UUCMS} scheme may be uninteresting from practical point of view as the source indices may be unknown in a lot of applications.}
\end{itemize}
%Please note that we can also think about the length function which intakes $x^n$ and its index $S$. It is straightforward to verify that the index $S$ does not provide any further useful information for the universal compression of $x^n$ and the performance of such a scheme is that of {\UUC}.
We quantify the performance of these fundamental schemes using their respective average redundancies.  
Let $R_\UC(l,\Delta)$ denote the average redundancy of the {\UUC} compression algorithm for the universal compression of a mixture source, which is defined in the usual way, as in~\eqref{eq:redundancy-def}, given by
\begin{equation}
R_\UC(l,\Delta) = \mb{E}l(X^n) - H_n(\Delta).
\end{equation}
Further, let $\underline{R}_\UC(n,\mb{w},\mb{d})$ and $\bar{R}_\UC(n,\mb{w},\mb{d})$   denote the average maximin and minimax redundancy, respectively, which are defined in the same manner as in~\eqref{eq:def-maximin} and~\eqref{eq:def-minimax} in Section~\ref{sec:background}. Our goal is to characterize the performance of universal compression as a function of the mixture weights $\mb{w}$ and source parameter vector dimensions $\mb{d}$. Note that the average maximin redundancies $\underline{R}_\UCM(n,m,\mb{w},\mb{d})$ and $\underline{R}_\UCMS(n,m,\mb{w},\mb{d})$, and the average minimax redundancies $\bar{R}_\UCM(n,m,\mb{w},\mb{d})$ and $\bar{R}_\UCMS(n,m,\mb{w},\mb{d})$ can also be defined similarly.

It is straightforward to extend Gallager's Theorem to the following.
\begin{theorem}
Consider {\UUC}, {\UUCM}, and {\UUCMS} for the compression of mixture sources with the set of parameter vectors $\Delta \in \Lambda'(n)$, where $\Lambda'(n)$ is defined in~\eqref{eq:lambda-prime}. Then, the average minimax redundancy and the average maximin redundancy are equivalent, i.e.,
\begin{align}
\bar{R}_\UC(n,\mb{w},\mb{d}) &= \underline{R}_\UC(n,\mb{w},\mb{d}) \nonumber\\
& = \max_{p}I(X^n;\Delta).\label{eq:channel:1}\\
\bar{R}_\UCM(n,m,\mb{w},\mb{d})&= \underline{R}_\UCM(n,m,\mb{w},\mb{d})\nonumber\\
& =  \max_{p}I(X^n;\Delta|\mb{Y}^{n,T}).\label{eq:channel:2}\\
\bar{R}_\UCMS(n,m,\mb{w},\mb{d}) &= \underline{R}_\UCMS(n,m,\mb{w},\mb{d}) \nonumber\\
&= \max_p I(X^n;\Delta| \mb{Y}^{n,T}, \mb{S},Z)\label{eq:channel:3}
\end{align}
Further, if $\Delta$ is chosen such that for $i \neq j$, we have $\theta^{(i)}$ and $\theta^{(j)}$ are independent and the marginal distribution of each $\theta^{(i)}$ is Jeffreys' prior on the $d_i$-dimensional space $\Lambda_{d_i}$, such prior is asymptotically capacity achieving as $n \to \infty$.
%$\Delta$ follows Jeffreys' prior.
\label{thm:Gallager2}
\end{theorem}
\begin{IEEEproof}
The proof is explained in the appendix.
\end{IEEEproof}
{\noindent \bf Remark.}
Note that our results hold for a set $\Lambda'(n) = \Lambda \setminus A(n)$ whose volume asymptotically equals that of $\Lambda$. In other words, if you pick the parameter vectors according to any distribution whose support is the entire set $\Lambda$ (i.e., it puts non-zero mass over any point in $\Lambda$), then our results would hold asymptotically almost surely (a.s.).\footnote{An event $A_n$ happens asymptotically almost surely (a.s.) if and only if $\lim_{n \to \infty} \mathbb{P}[A_n] =1$.}

Next, we state a trivial ordering on the average minimax redundancy of these {\em fundamental} schemes.
\begin{proposition}
\label{thm:order}
The following ordering holds for the average minimax redundancies of {\UUC}, {\UUCM}, and {\UUCMS}.
\begin{equation}
\bar{R}_\UCMS(n,m,\mb{w},\mb{d}) \leq 
\bar{R}_\UCM(n,m,\mb{w},\mb{d}) \leq 
\bar{R}_\UC(n,\mb{w},\mb{d}).
\end{equation}
\end{proposition}
\begin{IEEEproof}
This holds as the {\UUCMS} length function can choose to ignore $\mb{S}$ and $Z$, and also the {\UUCM} length function can choose to ignore $\mb{y}^{n,T}$. In other words, more information cannot hurt.
\end{IEEEproof}
In the rest of this section, our goal is to characterize the average minimax redundancies of the aforementioned fundamental schemes, and in particular the gaps between them, to understand the {\em fundamental} benefits provided by side information in the universal compression of a mixture of parametric sources.

\subsection{{\UUC}: Universal Compression without Side Information}
We refer to {\UUC} as the universal compression without side information, in which a universal length function $l(x^n)$ is used to compress the sequence $x^n$ without regard to the side information sequence $\mb{y}^{n,T}$.

Next, we state the main result in characterizing the average minimax redundancy.
%In what follows, we describe the finite-length limits of the universal compression on the mixture source.
\begin{theorem}
In the case of {\UUC}, we have
\begin{equation}
\bar{R}_\UC(n,\mb{w},\mb{d})= \sum_{d=1}^{d_\text{max}}v_d (\bar{R}_{n,d} -H_d) +  o(1) ~a.s.,
\end{equation}
\label{thm:UCOMP}
where $H_d$ is defined in~\eqref{eq:Hd}.
\end{theorem}
\begin{IEEEproof}
The proof is explained in the appendix.
\end{IEEEproof}
According to Theorem~\ref{thm:UCOMP}, in the universal compression of a sequence of length $n$ from the mixture source, %for sufficiently low-entropy $\mb{w}$ (or sufficiently small $\KK$),
the main term of the redundancy scales as the weighted average of $(\bar{R}_{n,d} - H_d)$ terms. This can be significantly large if $H(\mb{w}_d)$ is much smaller than $\frac{d}{2}\log n$.
Again, if $\KK=1$, we have $\bar{R}_\UC(n,1,d) = \bar{R}_{n,d}$; this is exactly the average minimax (maximin) redundancy in the case of one unknown $d$-dimensional source parameter vector described in Theorem~\ref{thm:maximin}.

Theorem~\ref{thm:UCOMP} also suggests that independently from $\KK$ and $H(\mb{w})$, the price to be paid for universality is given by $\bar{R}_{n,d}$ over and above $H_n(\Delta, Z)$, i.e., the entropy when $\Delta$ and $Z$ are known. In other words, $H(X^n) - H_n(\Delta,Z)$ scales like $\sum_d v_d\bar{R}_{n,d}$ (which is the price of universal compression of a sequence of length $n$ from a single source with an unknown $d$-dimensional parameter vector that follows the worst-case Jeffreys' prior averaged over $d$).

\begin{corollary}
If $ H(\mb{\hat{w}}_d)  \gtrsim \frac{d}{2}  \log n$ for all $1\leq d \leq d_\text{max}$, then
\begin{equation}
\bar{R}_\UC(n,\mb{w},\mb{d})= o(1) ~a.s.
\end{equation}
\end{corollary}
\begin{IEEEproof}
 If $v_d > 0$, then $H(\mb{\hat{w}}_d)  \gtrsim \frac{d}{2} \log n$, and hence, we have $H_d =  \bar{R}_{n,d}$, which means $\bar{R}_{n,d} - H_d$ vanishes. Hence, the main redundancy term $v_d (\bar{R}_{n,d} - H_d)$ in Theorem~\ref{thm:UCOMP} vanishes for all  $1\leq d\leq d_{\text{max}}$, which completes the proof.
\end{IEEEproof}
According to the corollary, for large $\KK$, we asymptotically almost surely (a.s.) expect no extra redundancy associated with universality on top of the mixture entropy. This is not surprising as even in the case of {\em known} source parameter vectors, as given by Theorem~\ref{thm:entropy_rate}, the redundancy converges to the weighted average of the redundancies for a  $d$-dimensional {\em unknown} source parameter vector that follow Jeffreys' prior. Therefore, there is no extra penalty when the source parameter vectors are indeed unknown. %In other words, the knowledge that $x^n$ is generated using one of so many sources is indeed useless.

\subsection{\UUCM: Universal Compression with Side Information}

We refer to {\UUCM} as the universal compression with side information. In this section, our goal is to characterize the average minimax redundancy of the {\UUCM} scheme given the side information, i.e., $\bar{R}_\UCM(n,m,\mb{w},\mb{d})$, where  $T=\frac{m}{n}$ sequences from the mixture source are shared between the encoder and the decoder as side information.

\begin{proposition}
\label{thm:small-m}
In the case of {\UUCM}, if $m = O(1)$, then
\begin{equation}
\bar{R}_\UCM(n,m,\mb{w},\mb{d}) = \bar{R}(n,\mb{w},\mb{d}) - O(1).
\end{equation}
\end{proposition}
According to Proposition~\ref{thm:small-m}, when $m$ does not grow to infinity, the improvement offered by side information is at most constant, which is negligible compared with the leading term of redundancy which is $O(\log n)$.

\begin{theorem}
In the case of {\UUCM}, for $m = \omega(1)$ we have
\begin{equation}
\bar{R}_\UCM(n,m,\mb{w},\mb{d})= \sum_{d=1}^{d_\text{max}}v_d\sum_{i=1}^\KK \hat{w}_{d,i} \hat{R}_{d,i} + o(1) ~a.s.,
\end{equation}
where $\hat{R}_{d,i}$ is given by
\begin{equation}
\hat{R}_{d,i}= \left\{
\hspace{-.05in}
\begin{array}{ll}
   \frac{d}{2}\log\left( 1 + \frac{n}{\hat{w}_{d,i} m}\right) +\delta& \text{if~} H(\mb{\hat{w}}_d) \lesssim \frac{d}{2} \log n \\
  0 & \text{if~} H(\mb{\hat{w}}_d) \gtrsim \frac{d}{2} \log n
 \end{array}
\right. ,
\label{eq:hat_R}
\end{equation}
where $\delta$ is an absolute constant with respect to $n$ and can be made arbitrarily small for sufficiently large $T$.
\label{thm:UCOM}
%\vspace{-.12in}
\end{theorem}
\begin{IEEEproof}
The proof is explained in the appendix.
\end{IEEEproof}
Theorem~\ref{thm:UCOM} characterizes the redundancy of the optimal universal compression scheme with side information, which uses a memory of size $m=nT$ ($T$ sequences of size $n$) in the compression of a new sequence of length $n$.
It is natural to expect that the side information will make the redundancy decrease. The redundancy of the {\UUCM} decreases when $H(\mb{w})$ or roughly $\KK$ is sufficiently small.
Again, $\KK=1$, results in $\bar{R}_\UCM(n,m,1,d) = \frac{d}{2} \log \left(1 + \frac{n}{m}\right) + o(1)$, which is consistent with what we derived for a single parametric source in~\cite{ISIT12_distributed}.
Further, it is deduced from Theorem~\ref{thm:UCOM} that $\lim_{T\to\infty} \bar{R}_\UCM(n,m,\mb{w},\mb{d})= o(1)$ (regardless of $\mb{w}$), i.e., the cost of universality would be negligible given that sufficiently large memory (side information) is available. Thus, the benefits of optimal universal compression with side information would be substantial when $H(\mb{w})$ is sufficiently small. On the other hand, when $H(\mb{w})$ grows very large, no benefit is obtained from the side information in the universal compression and the performance improvement becomes negligible. This is due to the fact that,  in light of Theorem~\ref{thm:UCOMP}, the compression performance for the known source parameters case is already equal to that of the universal compression.

\subsection{\UUCMS}
Next, we analyze the fundamental performance of a class of schemes that have access to the unknown source labels. In particular, we would like to analyze how much performance improvement the knowledge of the unknown source indices would offer over the fundamental limits of {\UUCM}.
We refer to {\UUCMS} as the universal compression with perfectly clustered side information sequence $\mb{y}^{n,T}$, which  is shared between the encoder $E$ and the decoder $D$. 
%Further, it is assumed that $E$ and $D$ have access to an oracle that can determine the index of the source that generated each sequence. 
Further, the index vector $\mb{S}$ of the memorized sequences and the index $Z$ of the sequence $x^n$ to be compressed are known to both $E$ and $D$. Therefore, one can imagine that an oracle exists that can partition the sequences in $\mb{y}^{n,T}$ based on their source index. Then, it can be shown that it is optimal that $E$ and $D$ cluster the side information sequences according to $\mb{S}$ and use the minimax estimator to estimate the source parameter vector associated with each cluster; the encoder $E$ classifies the sequence $x^n$ to the respective cluster using the oracle and encodes the sequence only using the side information provided by the estimated parameter vector of the respective cluster.

\begin{theorem}
\label{thm:UCPCM}
In the case of {\UUCMS}, we have
\begin{equation}
\bar{R}_\UCMS(n,m,\mb{w},\mb{d})= \sum_{d=1}^{d_\text{max}}v_d\sum_{i=1}^\KK \hat{w}_{d,i} \hat{R}_{d,i}+ o(1)~ a.s.,
\end{equation}
where $\hat{R}_{d,i}$  is defined in~\eqref{eq:hat_R}.
\end{theorem}
\begin{IEEEproof}
The proof is explained in the appendix.
\end{IEEEproof}
Theorem~\ref{thm:UCPCM} characterizes the redundancy of the universal compression with perfectly clustered side information. It is straightforward to observe that for sufficiently large $m$, the redundancy of {\UUCMS} becomes very small.
However, {\UUCMS} is impractical in most situations as the oracle that provides the source index is not available. %Furthermore, {\UUCMS} is also not necessarily optimal for all $n$. 
As an important special case if $\KK=1$, then $\bar{R}_\UCMS(n,m,\mb{w},\mb{d}) = \frac{d}{2} \log \left(1 + \frac{n}{m}\right) + o(1)$, which reduces to Theorem 2 of~\cite{ISIT12_distributed} regarding the average minimax redundancy for the case of a single source with an unknown parameter vector.

\begin{corollary}
Regardless of $\mb{w}$ and $\mb{d}$, we have
\begin{equation}
\lim_{T \to \infty} \bar{R}_\UCMS(n,m,\mb{w},\mb{d}) = o(1).
\end{equation}
\end{corollary}
\begin{IEEEproof}
Note that $T\to \infty$ simply means $m \to \infty$, and $\frac{d}{2}\log\left( 1 + \frac{n}{\hat{w}_{d,i} m}\right) \to 0$ as $m \to \infty$, completing the proof.
\end{IEEEproof}
According to the corollary, the redundancy vanishes as $T \to \infty$ (or equivalently $m \to \infty$). Therefore, for sufficiently large $m$, significant performance improvement is expected in terms of the number of bits required to describe a sequence $x^n$.

\begin{corollary}
We have
\begin{equation}
\bar{R}_\UCM(n,m,\mb{w},\mb{d})= \bar{R}_\UCMS(n,m,\mb{w},\mb{d}) + o(1) ~a.s.
\end{equation}
%\vspace{-.12in}
\end{corollary}
\begin{IEEEproof}
The corollary is proved by combining Theorems~\ref{thm:UCOM} and~\ref{thm:UCPCM}.
\end{IEEEproof}
\vspace{0.12in}
{\noindent \bf Remark.}
The corollary has significant implications. It states that the performance of optimal  universal compression with side information  ({\UUCM}), which uses a memory of size $m=nT$ ($T$ sequences of size $n$) in the compression of a new sequence of length $n$ is equal to that of the universal compression with perfectly clustered memory ({\UUCMS}) up to $o(1)$ terms.
Hence, when $T$ is sufficiently large, we expect that both have the same performance. This indeed demonstrates that {\em clustering} is optimal for the universal compression with side information. As such, we pursue the clustering of the side information (i.e., memory) in this paper in Section~\ref{sec:cluster}.

\section{Operational Limits of Universal Compression for Mixture Sources}

In addition to the fundamental schemes (and respective length functions), in this paper, we will also assess two {\em operational} schemes listed below. Both schemes fall in the {\UUCM} coding regime that we have access to the memory but not the source indices.
\begin{itemize}
\item {\UUCMI}: Simple universal compression with side information (common memory between the encoder and the decoder), which treats the side information as if it were generated from a single parametric source. In other words, it uses the minimax estimator for the unknown parameter vector of the source for a single source. The length function associated with this operational scheme is denoted by $l_\text{M}^1(x^n, \mb{y}^{n,T})$.
\item {\UUCMc}: Universal compression with clustering of the side information, which is the practical clustering-based scheme proposed in this paper and shall be described in Section~\ref{sec:cluster}.
\end{itemize}
Since for these operational schemes the length function is predetermined, we will quantify their performance under the worst-case prior on the space of the source parameter vectors. The worst-case prior is derived as a by-product of Theorem~\ref{thm:Gallager2}.

\subsection{{\UUCMI}: Simple Universal Compression with Side Information}
Next, we comment on the performance of the simple universal compression with side information scheme that is regarded as {\UUCMI}. In this compression scheme, it is assumed that the encoder $E$ and the decoder $D$ (in Fig.~\ref{fig:single_hop}) both have access to the memorized sequence $\mb{y}^{n,T}$ from the mixture source.  The sequence $\mb{y}^{n,T}$ is used to form the optimal minimax estimator of one unknown source parameter vector. Observe that the scheme would be minimax optimal if $\mb{y}^{n,T}$ was generated by a single parametric source with an unknown parameter vector. The estimated source parameter using the minimax estimator for one unknown parameter vector is then used for the compression of the sequence $x^n$.

As discussed in Section~\ref{sec:entropy}, when the source parameter vectors are known, then mixing the probability distributions is optimal and achieves the entropy. The subtlety here is that since the source parameter vectors are unknown, there is a penalty to be paid for learning them.
When the source is a mixture of more than one source parameter vectors, {\UUCMI} will naively start to build a larger model for the source with much more parameters for it to be able to closely follow the source statistics. As the length of the sequences become sufficiently large, such an approach will be able to learn the source statistics fairly well. It will indeed converge to the source model as the depth of the built context tree grows but with significantly larger number of parameters. Unfortunately, model reduction methods such as context pruning~\cite{Baron_O_n} will not be a remedy to this issue either. We will comment more on the performance of {\UUCMI} in Section~\ref{sec:simulation}.

\subsection{{\UUCMc}: Universal Compression with Clustering of the Side Information}
Thus far, we argued why a naive memory-assisted compression ({\UUCMI}) would suffer from curse of dimensionality in learning the unknown source parameters. On the other hand, in Section~\ref{sec:universal}, we theoretically proved that the optimal memory-assisted compression performs similarly to the memory-assisted compression with known source indices.  This suggests that an asymptotically optimal strategy would be to cluster the side information sequences into several distinct models (one model for each source in the mixture).
This shall significantly reduce the number of parameter models and hence will improve the compression performance. This is the subject of the next section of this paper.

%%%%%%%%%%%%%%%%%%%%%%%%%%%%%%%%%%%%%%%%%%%%%%%%%%%%%%
\section{Clustering Algorithms for Compression of Mixture Sources}
\label{sec:cluster}
\begin{figure}
\begin{center}
  \includegraphics[width=\linewidth]{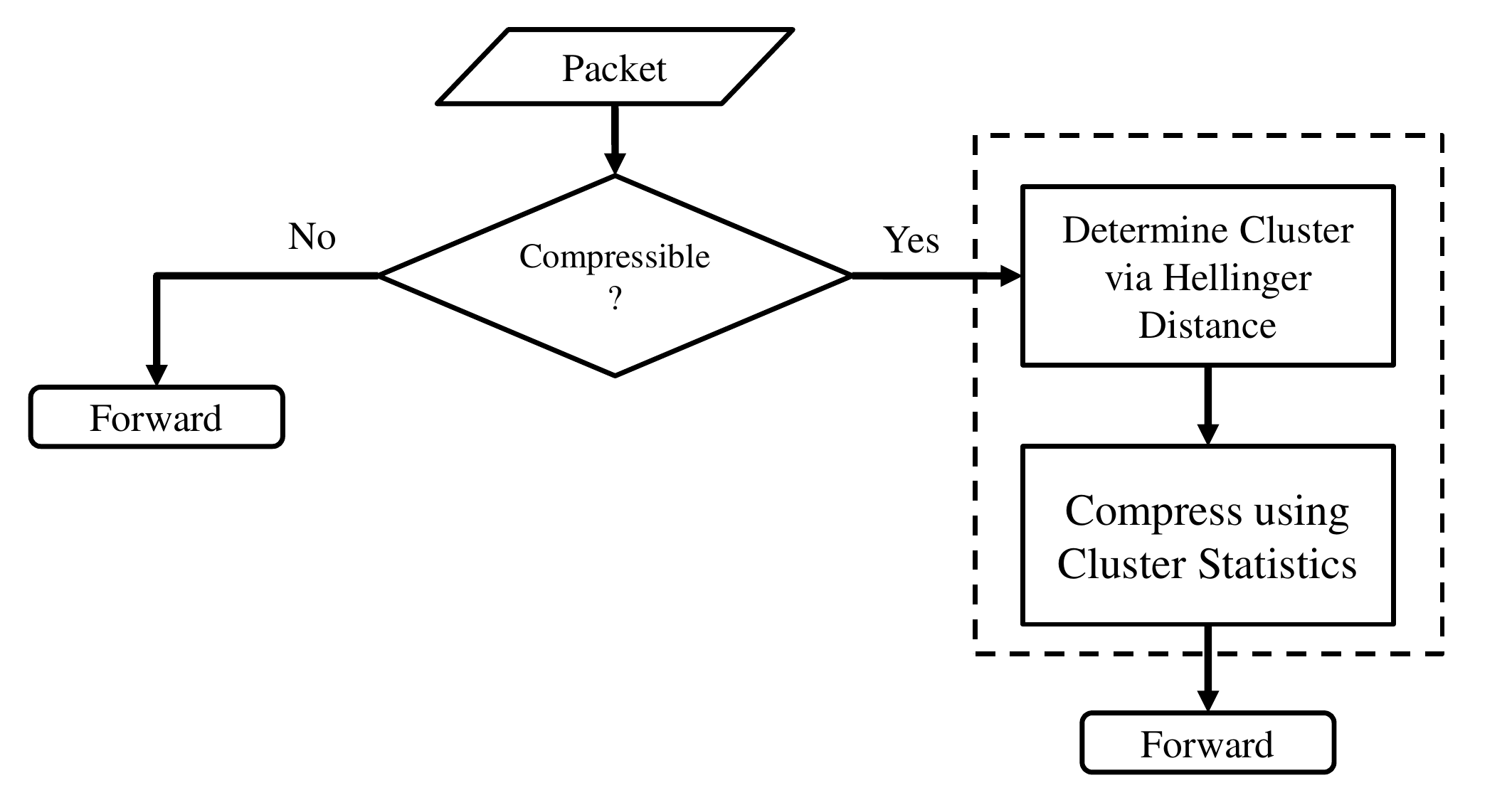}
%  \vspace{-.1in}
  \caption{Network packet compression flowchart. The modules in the dashed box are the components of the k-means clustering using Hellinger distance.}
  \label{fig:flowchart}
\end{center}
%\vspace{-.3in}
\end{figure}

In this section, we present two clustering solutions for network packets. The k-means algorithm can be used for this purpose provided that a proper feature space and a relevant distance metric are selected. Further, we have also experimented with the non-parametric k-nearest neighbors clustering algorithm and we will comment on the performance of both algorithms.
In the sequel, we describe a hierarchical clustering algorithm that proves to be useful for compression.
The proposed hierarchy for the content-aware joint memorization and clustering for network packet compression is shown in Fig.~\ref{fig:flowchart}. As shown, we first identify whether or not an incoming packet is compressible. If the packet is determined incompressible, it is neither compressed nor stored in the memory. On the other hand, the compressible packets are passed to the clustering unit which operates based on the Hellinger distance metric.

%The k-means clustering algorithm~\cite{Bishop2006} requires the number of clusters to be known apriori which is not the case in general. To address this limitation, principled clustering algorithm are created that does not need to know the number of clusters in advance.
%%%%%%%%%%%%%%%%%%%%%%%%%%%%%%%%%%%%%%%

{\bf Compressibility determination:}
The compressibility determination is performed based on the empirical entropy of the data packet.
The side information packets in memory may be divided into two categories: one category contains packets with very high entropy rate (close to 8 bits per byte) and hence these packets are incompressible. These include already compressed videos or images. The other category contains packets whose empirical entropy rate is estimated to be much less than 8, and hence, these packets are compressible. Therefore, as the first step, the packets are partitioned into compressible and incompressible. 
%The algorithm for compressibility determination is demonstrated in Alg.~\ref{algo:compressibility}.
After the partitioning step, the packets in the resulting memory are all compressible. Then,  we will perform a clustering of the resulting memory.
Note that one may generate man-made models where each sequence has high entropy while the individual sequences are indeed highly correlated. On the other hand, our observations from the real data traces suggest that this issue is not encountered in practice. Hence, we chose to ignore the packets that we determine to be incompressible.

{\bf Feature selection:}
Feature extraction deals with extracting simpler descriptions for a large set of data that can accurately describe characteristics of original data. For memoryless source models, the frequency of each alphabet in the sequence defines an empirical probability density function (pdf) vector which also happens to be the sufficient statistics.
Although for more sophisticated source models, the empirical pdf of the packet (i.e., the frequency of each byte in the packet) is not a sufficient statistics anymore as collisions may occur between different parametric sources in the marginal symbol distribution, the empirical probability distribution would still match for packets from the same parametric source while the probability of collision is relatively low. Further, since the lengths of the data packets are relatively small on the order of several kilobytes, any model beyond a memoryless model would overfit the data~\cite{Csiszar06, optimal_Markov_order,kieffer_order,Barron_Cover_91}. Hence, we assume that each packet is generated using a memoryless model. We choose the vector of the empirical pdf as our feature vector and since we work at the byte granularity (i.e., $|\mc{X}| = 256$), the feature vector is $255$-dimensional (255 independent variables). We stress that the chosen feature space is not necessary optimal but simulations confirm that it works close to optimal in practice for packets of size 1,500 bytes or longer.

{\bf Distance metric:}
To perform clustering, we need to use a distance metric that determines the similarity between any two packets. Note that the overall objective is to reduce the compression rate where the compression penalty can be described in terms of KL-divergence between the true model and the estimated model (cf.~\cite{Barron_Cover_91}). On the other hand, KL-divergence is not a metric. Hence, the natural choice for the distance metric would be the Hellinger distance metric, which is widely used to quantify the similarity between two probability distributions (cf.~\cite{Hellinger}).
%The Hellinger distance is a metric ~(cf.\cite{Hellinger}).
For two probability distributions $p(\cdot)$ and $q(\cdot)$ defined on symbols from alphabet $\mc{X}$, the Hellinger distance is defined as
%\vspace{-.1in}
\begin{equation}
d_H(p, q) = \frac{1}{2}\sqrt{\sum_{x\in \mc{X}}\left(\sqrt{p(x)}-\sqrt{q(x)}\right)^{2}}.
\vspace{-.05in}
\end{equation}
In our setup, we calculate the Hellinger distance of two packets using the empirical pdf of the symbols for each packet.
Recall that a packet $x^n \in \mc{X}^n$ is a vector of $n$ symbols $x_i \in \mathcal{A}$.

%%%%%%%%%%%%%%%%%%%%%%%%%%%%%%%%%%%%%%%%%%%%%%%%%%%%%%%%%%%%%
{\bf k-means clustering:}
As discussed earlier in Section~\ref{sec:problem_setup}, we have a side information sequence of packets $\mb{y}^{n,T}$ that consists of $T$ packets that originated from a mixture source model.
We stress that the total number of source in the mixture (denoted by $K$) is unknown. Each packet in the memory needs to be assigned to one clusters from the $k$ choices. We use the binary indicator $c^j_{t}$ to denote the cluster assignment for the $t$-th packet $y^n(t)$. The indicator $c^j_{t} = 1$ if $y^n(t)$ is assigned to cluster $j\in[k]$, otherwise $c^j_{t} = 0$. Then, the objective function for  clustering is given by
\vspace{-.1in}
\begin{equation}
\label{eqn: cluster_obj}
J = \sum_{t = 1}^{T}\sum_{j = 1}^{k}c^j_{t}d_H(q_{t},u_{j}),
\vspace{-.08in}
\end{equation}
where $q_{t}$ is the distribution on the symbols obtained from $y^n(t)$  and $u_{j}$ is the probability distribution vector on the symbols associated with the packets in cluster $j$. The goal of the clustering algorithm is to find the assignment $c^j_{t}$ for $j \in [k]$ and $t \in [T]$ such that $J$ is minimized.

The problem setup suggests that the k-means clustering algorithm~\cite{Bishop2006} is suitable for our purpose. k-means algorithm is an iterative algorithm which consists of two steps for successive optimization of $c^j_{t}$ (and hence $u_{j}$). 
Given cluster center $u_{j}$, the optimal $c^j_{t}$ can be easily determined by assigning the packet $y^n(t)$ to the closest cluster with minimum Hellinger distance $d_H(q_{t}, u_{j})$. Then, we fix $c^j_{t}$ and update $u_{j}$.
k-means clustering algorithm can successfully cluster data packets in the ideal situation with static number of source model mixture. However, this algorithm can break down when the number of sources cannot be estimated correctly, especially for the infinite mixture source model in real world networks. Note that k-means algorithm also requires the selection of k a priori. In our simulations we observed that choosing a large k would always do the job as most of the clusters will remain empty when the algorithm converges.

\begin{figure}
\centering
%\vspace{0.1in}
\includegraphics[width = \linewidth]{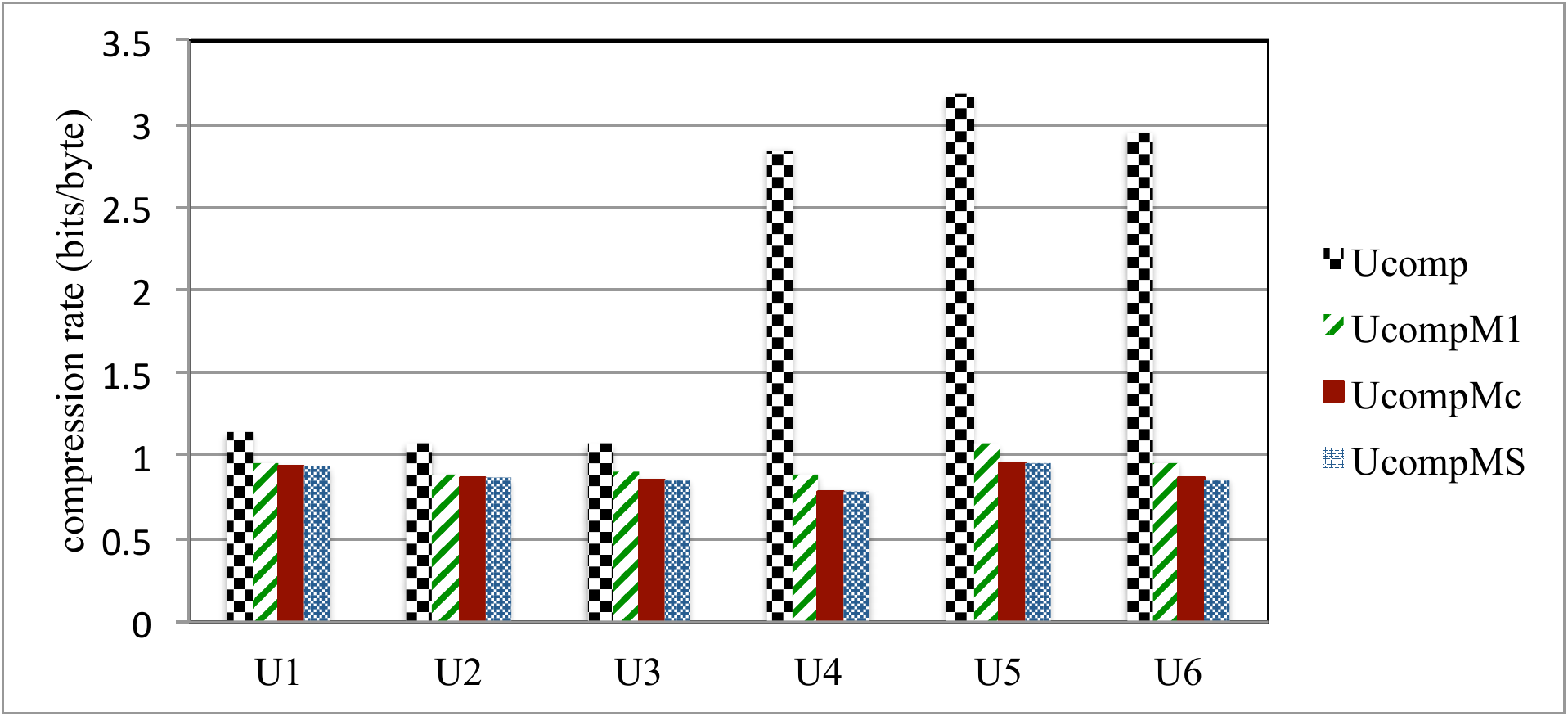}
%\vspace{0.1in}
\caption{Average compression-rate for a mixture of 3 memoryless and 3 first-order Markov sources using Lite PAQ compression algorithm.}
%\vspace{-.1in}
\label{fig:man-made}
\end{figure}

{\bf Non-parametric k-nearest clustering:}
To cluster packets without assuming any parameters a priori about the data, we also used the dynamic non-parametric clustering method based on the well known $k$-nearest algorithm. To this end, we partition the memory into $m$ small sub-clusters that are represented by the cluster centers $S=\left\{s_{1},\ldots, s_{m}\right\}$. Each sub-cluster consists of about $T/m$ neighboring packets with the minimum variance.

As soon as the fine-grain sub-clusters are produced, then we can process the training packets to form the appropriate memory for compression. %Each sub-cluster is represented by the mean value of the included vectors, the similarity between sample packet $x^{n}$ and sub-clusters is measured by the distance between $x^{n}$ and $s_{i}$.
 After the initialization of the current sub-cluster set $C=S$, the sub-cluster from set $C$ nearest to $x^{n}$ is merged into the training set $Q$ and is removed from $C$ after merging. In other words, the new dynamic training set $Q$ is updated. The merging ends when the expected number of training packets is reached. The actual number of sub-clusters is fixed according to the minimum number of packets requirement of compressor. Algorithm \ref{alg:clustering} elaborates the procedures of the non-parametric clustering for selection of training packets.
\begin{algorithm}[h]
  \caption{Non-Parametric $k$-Nearest Clustering Algorithm}
  \label{alg:clustering}
  \begin{algorithmic}
%    \State Compute empirical PDF vectors $\left\{d_{i}\right\}$
    \State Compute sub-cluster centers $S=\left\{s_{1},\ldots, s_{m}\right\}$
    \For{ Incoming packet $x^{n}$}
    \State Compute distance $d_H(x^{n},s_{i})$
    \State Current sub-cluster set $C=S$
    \While{ $training\_pkt\_num\!\!<\!\!min\_training\_num$}
    \If {$s_{i^\star}=\min_{s_{i}\in C}d_H(x^{n},s_{i})$}
    \State Training set $Q=Q\cup \left\{s_{{i}^\star}\right\}$
    \State Index set $T=T\cup \left\{{i}^\star\right\}$
    \State $training\_pkt\_num$ update
    \State Remove $s_{{i}^\star}$ from $C=\left\{s_{1},\ldots, s_{m}\right\}$
    \EndIf
    \EndWhile
    \State Return $Q$ and $T$
    \EndFor
  \end{algorithmic}
\end{algorithm}

In practice,  the feature vectors of data packets are scattered in a high dimensional space and the shapes of clusters are arbitrary. In particular, when the sample data packet does not belong to any of the clusters, the performance of k-means clustering will be adversely impacted. %The clustering algorithm take this characteristic of real world data as the basic assumption. 
On the other hand, by merging nearby sub-clusters, k-nearest algorithm can collect most useful training data with appropriate consistency for sample packet compression. Besides, without the knowledge of the number of clusters in advance, the k-nearest clustering algorithm achieves performance improvement compared to k-means clustering. All the detailed simulation in next session will elaborate on the performance of the k-nearest algorithm for data compression.

\begin{figure*}
\centering
%\vspace{0.1in}
\includegraphics[width = 0.8\linewidth]{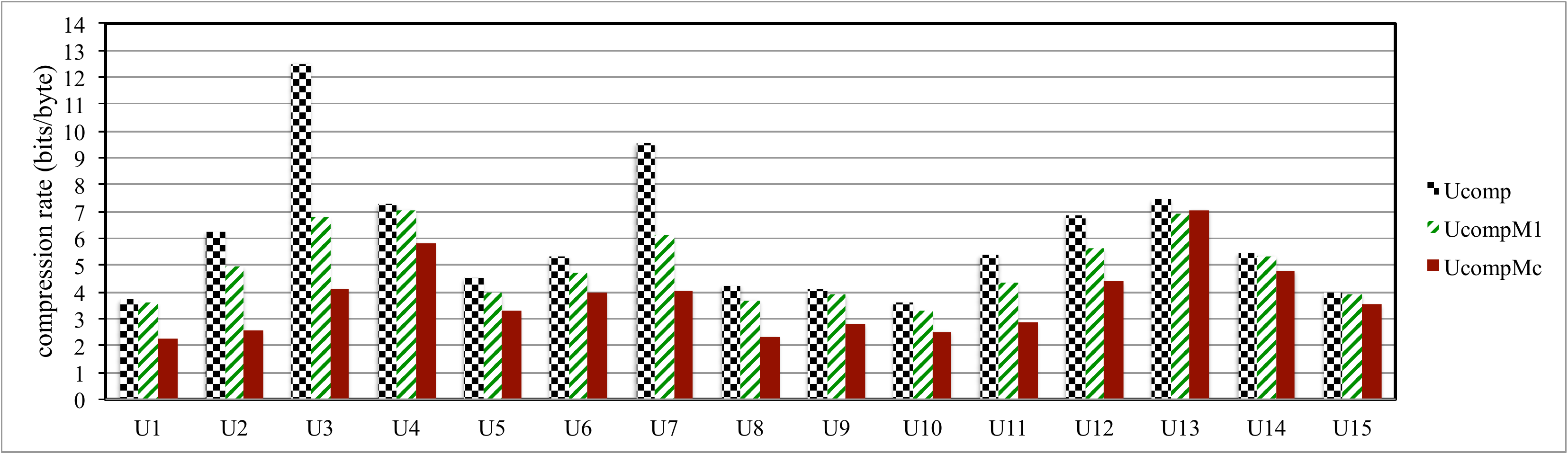}
%\vspace{0.1in}
\caption{Average compression-rate of GZIP on real traffic data.}
%\vspace{-.1in}
\label{fig:real-data-gzip}
\end{figure*}

\begin{figure*}
\centering
%\vspace{0.1in}
\
\includegraphics[width = 0.8\linewidth]{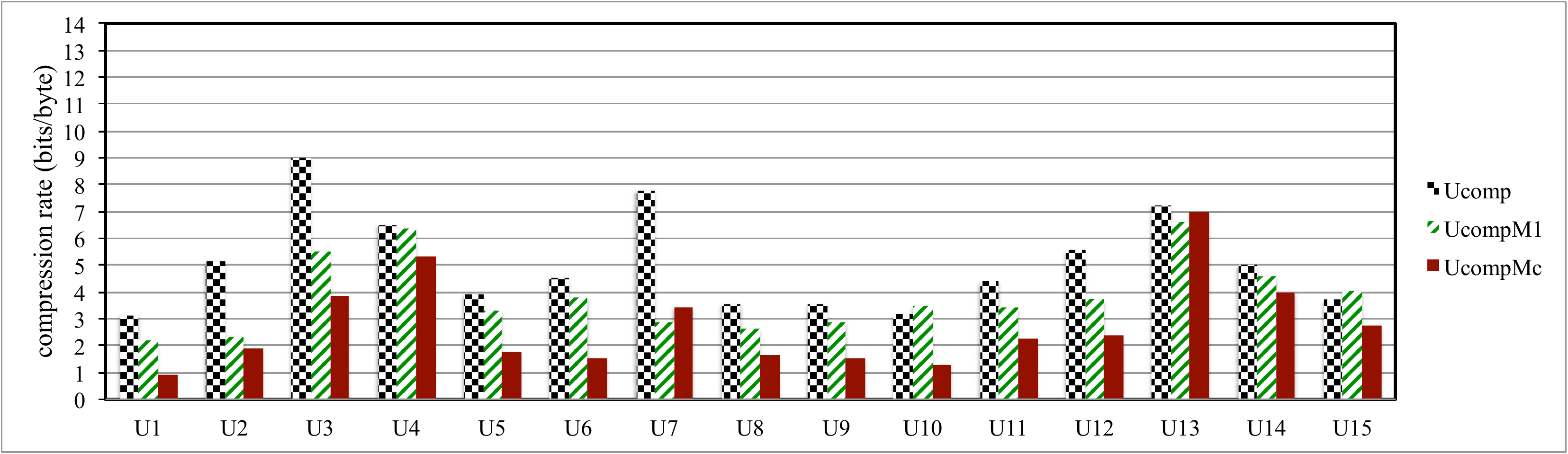}
%\vspace{0.1in}
\caption{Average compression-rate of CTW on real traffic data.}
%\vspace{-.1in}
\label{fig:real-data-ctw}
\end{figure*}

\begin{figure*}
\centering
%\vspace{0.1in}
\
\includegraphics[width = 0.8\linewidth]{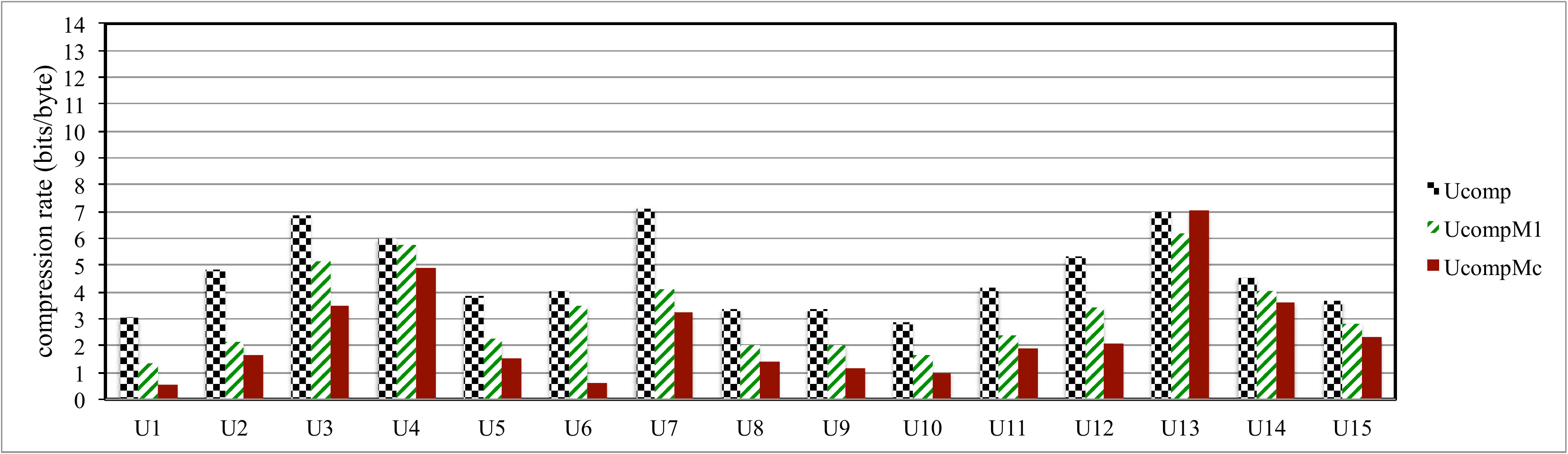}
%\vspace{0.1in}
\caption{Average compression-rate of Lite PAQ on real traffic data.}
%\vspace{-.1in}
\label{fig:real-data-lpaq}
\end{figure*}

%%%%%%%%%%%%%%%%%%%%%%%%%%%%%%%%%%%%%%%%%%%%%%%%%%%%%%%%%%%%%
{\bf Compression of a new packet:}
Once the clustering of memory is performed, we will derive the mixture distribution by mixing all the distributions obtained from the source models, as discussed in Section~\ref{sec:entropy}, to achieve the source model. Then, the new sequence can be compressed on the fly without any further processing. This is a perfect fit for the statistical compression methods, such as the CTW and LPAQ.

For dictionary based methods, since mixing is impossible, we perform classification 
to compress a new packet $x^n$. We
first decide which cluster should be used as the side information to compress $x^n$. Therefore, we classify the packet $x^n$ by assigning it to a proper cluster.
The classification algorithm is as follows. Let $c$ be the cluster label of $x^n$ to be determined. We compute Hellinger distance between the symbol distribution $q$ of $x^n$ and the cluster $u_{j}$. Then $x^n$ is assigned to the closest cluster by
\vspace{-.08in}
\begin{equation}
c = \underset{1 \le j \le K}{\text{argmin}} \, d_H(q, u_{j}).
\vspace{-.05in}
\end{equation}
%The classification algorithm is summarized in Alg.~\ref{algo:classification}.

\section{Simulation and Evaluation}
\label{sec:simulation}
In this section, we present simulation results to demonstrate the performance of the proposed memory-assisted compression system with non-parametric clustering and the overall improvement obtained from side information in universal compression of a mixture of parametric sources. Furthermore, we discuss the trade-off between compression speed and performance.

\subsection{Simulations on Man-Made Mixture Models}
To validate the theoretical results of the paper, we chose to use a mixture of parametric sources as the content-generator for the traffic. In particular, we used a mixture of five memoryless and five first-order Markov sources on 256-ary alphabet ($\left|\mc{X}\right|=256$). Consequently for a memoryless source the number of source parameters $d = 255$, while for a first-order Markov source $d$ is $256\times 255$ which is the number of independent transition probabilities. Further, we assume that each packet is selected uniformly at random from the above mentioned mixture. For short-length sequences, we generate 18,000 packets at random from this source model, where each packet is 1,500 bytes long. Then, we used 200 packets from each source as test packets for the purpose of evaluation.

Fig.~\ref{fig:man-made} demonstrates the results of the simulation on man-made data generated from the described mixture source using Lite PAQ compression algorithm. This plot shows the compression rate measured in the number of bits required to describe each source byte. Hence, the uncompressed source would need 8 bits/byte. Sources U1 through U3 are memoryless whilst sources U4 through U6 are first-order Markov sources. As can be seen, when the source model is simpler universal compression (without side information can work relatively much better and get closer to the entropy) whereas when the source model is first-order Markov there is a 300\% gap between the performance of the universal compression without side information and with side information. Further, as can be seen, the benefits of {\UUCMc} over {\UUCMI} become more spelled out when the source model becomes more complex. We will see simulations on real data in the next section.

\subsection{Simulations on Real Network Traces}
For a realistic evaluation, we perform simulation with data gathered from 20 different mobile users network traces in real world. The data set was gathered by Sanadhya {\em et al.} in~\cite{Shruti2012}. First, we randomly generate packet sequences from the 27,000-packet mixture of 15 users to construct the commonly accessible memory for clustering. Then, 10 sample packets from each of the 20 users (200 packets in total) are selected as test packets. Note the test packets are distinct from the packets used for training. Besides, there are 50 test packets that are generated from the 5 users which are not used for the generation of the training packets and hence do not have packets in the mixture memory. Average compression rate of each test packet is taken as the compression performance metric. We stress that each test packet is compressed separately and the result is averaged over the sample test packets. This is due to the packets flow in networks is a combination of packets from different sources and can not be simply compressed together. The whole simulation setup is summarized in Table~\ref{tab:sim_setup}.

\begin{table}
\caption{Simulation Setup Summary} % title of Table
\centering % used for centering table
\begin{tabular}{c c } % centered columns (4 columns)
\hline\hline %inserts double horizontal lines
Case &  Value \\ [0.5ex] % inserts table
%heading
\hline % inserts single horizontal line
No. of users in mixture source & 15 \\
No. of packets from each user in memory & 1,800\\
Total no. of memory data packets & 27,000 \\
Average size of each data packet & 1kB \\
Approximate size of total memory & 25MB \\
No. of users for performance testing & 15 \\
Total number of test packets & 200 \\
Distance metric & Hellinger distance \\
Clustering algorithm & k-means, k-nearest\\
Compression algorithm &  Gzip, CTW, Lite PAQ\\
 [1ex] % [1ex] adds vertical space
\hline %inserts single line
\end{tabular}
\label{tab:sim_setup} % is used to refer this table in the text
\end{table}

To demonstrate the impact of the side information on the compression performance, we analyze the average compression rate of the three important schemes (Ucomp, UcompM1, and UcompMc) using gzip, CTW, and Lite PAQ in Figs.~\ref{fig:real-data-gzip}, \ref{fig:real-data-ctw}, and~\ref{fig:real-data-lpaq}, respectively. Please see~\cite{TON14} for a discussion on the pros and cons of using each of these compression algorithms for network data compression.
 As can be seen, universal compression without help of any memory packets ({\UUC}) results in the largest (worst) compression-rate which verifies the penalty of finite-length compression analyzed in~\cite{ISIT11}. {\UUCMc}, which is the cluster-based memory-assisted compression, consistently outperforms all other schemes. It is worth noting that for the data from users which are not necessarily from mixture source model (users T1,\ldots, T5), non-parametric clustering still achieves impressive improvement compared to simple memory assisted compression {\UUCMI}. Compression with memory of user's previous packets {\UUCMS} sometimes performs well while it sometimes performs poorly due to the fact that the user data possibly comes from variant source models. In general, clustering algorithm is applicable to both Lite PAQ compression and CTW compression with impressive improvement.

 Table~\ref{tab:average} presents the average traffic reduction over all the fifteen users with different compression algorithms. Using the non-parametric clustering scheme, we compare the overall improvement of both dictionary-based compressor (Gzip)~\cite{gzip} and statistical compressor (Lite PAQ and CTW). As can be seen, Lite PAQ (which is close to the state-of-the art in  compression) achieves nearly 70\% traffic reduction and CTW achieves 65\% reduction. With more than 65\% traffic reduction, statistical compression outperforms dictionary-based compression, which offers 60\% reduction. However, dictionary-based compression tends to have ten times higher compression speed. Wireless applications tolerate more latency compared to the wired networks. Hence, statistical compression is more suitable for wireless data compression while dictionary-based compression is likely to be employed in wired networks.

\begin{table}[t]
\centering
\caption[]{The average compression rate  (bits/byte) of different compression schemes on the real network traffic traces.}
%\vspace{-.11in}
\begin{tabular}{|l|c|c|c|c|}
\hline
~ 	&\!\!{\UUC}\! &\! {\UUCMI}\! &\! {\UUCMc}  \\

\hline
Gzip 			    	& 6.01	 &		4.95 &	3.14 \\
\hline
CTW 			& 5.10	& 3.85	&      2.77 \\
\hline
Lite PAQ			& 4.66	 	&	3.25 &	2.43 \\
\hline
\end{tabular}
\label{tab:average}
\vspace{-.08in}
\end{table}

\subsection{Clustering Algorithm Performance Comparison}
%To compare the overall performance of memory-assisted compression with different clustering schemes, simulation results are summarized in Table~\ref{tab:setup}.

We choose packet selection with two algorithms, namely, k-means clustering algorithm and $k$-nearest clustering algorithm. According to Table~\ref{tab:setup}, non-parametric clustering achieve very similar performance, around 8\% better than k-means clustering. Besides, non-parametric clustering does not require to know the number of clusters in advance like k-means clustering. By using ball tree data structure~\cite{balltree}, the computational cost of nearest sub-clusters search is $O(N\log(N))$, where $N$ is the number of sub-clusters. The average size of training packets selected by k-means clustering is around 1800 packets whereas around 200 packets by non-parametric clustering. With smaller-sized training packet selected by $k$-nearest clustering algorithm, the compression speed is 9 times quicker than that of k-means clustering. As the average size of clusters generated from k-means is 9 times larger than the non-parametric counterpart.
 Through compression performance, the $k$-nearest clustering algorithm is proved to be more effective in network traffic redundancy reduction than referenced k-means clustering algorithm for real world data.

\begin{table}[t]
\centering
\caption[]{Average Compression Rate (bits/byte) of {\UUCMc}  for  Different Clustering Schemes and Compression Algorithms.}
%\vspace{-.11in}
\begin{tabular}{|l|c|c|c|}

\hline
{\UUCMc}	 & 	k-means 	& 	k-nearest \\
\hline
Gzip 				 	& 	 3.75	& 	3.14  \\
\hline
CTW 				 	& 	3.02	& 	2.77  \\
\hline
Lite PAQ					& 	2.63 	& 	2.43 \\
\hline
\end{tabular}
\label{tab:setup}
%\vspace{-.28in}
\end{table}
%\emph{3) Comparison of compression algorithms}: With the best clustering scheme from the above two comparison, we compress the overall improvement of both dictionary-based compressor (Gzip) and statistical compressor (Lite PAQ and CTW). We discuss of trade-off between the performance and compression speed in performance evaluation.

\section{Conclusion}
\label{sec:conclusion}
In this paper, we derived the fundamental limits of universal compression (with and without side information) for  mixture sources. Our results showed that significant improvement can be expected from side information in the universal compression of mixture sources. Our results further demonstrate that the optimal performance using side information corresponds to that of universal compression with known source indices. Motivated by this result, we presented two clustering algorithms for the universal compression of mixture sources with side information and demonstrated their effectiveness on data gathered from real network traces.

\ifCLASSOPTIONcaptionsoff
  \newpage
\fi

\appendix

\begin{IEEEproof}[Proof of Theorem~\ref{thm:entropy_rate}]
Let $D$ be the random dimension of the source parameter vector. It is straightforward to show that
\begin{equation}
H(X^n|\Delta) = H(X^n|\Delta,Z, D) + I(X^n;Z,D|\Delta)
\end{equation}
Further, if $Z$ is known, $D$ is determined, and hence, $H(X^n|\Delta,Z,D) = H(X^n|\Delta,Z)$ which is derived in~\eqref{eq:base_entropy}.
On the other hand, we have
\begin{equation}
I(X^n;Z,D|\Delta) = I(X^n;D|\Delta) + I(X^n;Z|\Delta, D).
\end{equation}
Let us first focus on $I(X^n;D|\Delta)$. We have
\begin{equation}
I(X^n;D|\Delta) = H(D|\Delta) - H(D|\Delta, X^n).
\end{equation}
Note that $H(D|\Delta)$ is by definition equal to $H(\mb{v})$. Further, we can use the maximum likelihood estimator of $D$ using $x^n$,  asymptotically as $n \to \infty$, to consistently estimate $D$ asymptotically almost surely~\cite{Csiszar06}.\footnote{An event $A$ happens a.s. (almost surely) if and only if $\mathbb{P}[A] =1$.} Hence, $ H(D|\Delta, X^n) = o(1)$  and $I(X^n;D|\Delta) = H(\mb{v}) + o(1)$.

Next, we consider $I(X^n;Z|\Delta,D)$. In this case, we have
\begin{equation}
I(X^n;Z|\Delta,D) = \sum_{d=1}^{d_\text{max}} v_d I(X^n;Z|\Delta,D=d).
\end{equation}
In order to analyze, we need to consider two situations. First, let $H(\mb{\hat{w}}_d) \lesssim \frac{d}{2} \log n$. We have
\begin{align}
I(X^n;Z|\Delta,D=d) &= H(Z|\Delta,D=d)\nonumber \\
&- H(Z|X^n,\Delta,D=d).
\end{align}
Clearly, $H(Z|\Delta, D=d) = H(\mb{\hat{w}}_d)$ by definition. Furthermore, the maximum likelihood estimator for the source parameter vector almost surely converges to the true $\theta$ in mean square with variance $O\left(\frac{1}{n}\right)$. On the other hand,  if $H(\mb{\hat{w}}_d) \lesssim \frac{d}{2} \log n$, let $A(n)$ contain all $\Delta$ where there exist two parameter vectors such that $||\theta^{(i)} - \theta^{(j)}|| = O\left(\frac{1}{\sqrt{n}}\right)$. It is straightforward to see that the volume of such set shrinks to zero as $n \to\infty$. Now, we only consider the set $\Lambda'(n)$ defined as 
\begin{equation}
\Lambda'(n) = \Lambda \setminus A(n).
\label{eq:lambda-prime}
\end{equation}
Then, for $\Delta \in \Lambda'(n)$, we have for any parameter vector $\theta^{(i)} \in \Delta$, all other parameter vectors are asymptotically such that $||\theta^{(i)} - \theta^{(j)}|| = \omega\left(\frac{1}{\sqrt{n}}\right)$. Hence, by picking the closest parameter vector to the maximum likelihood estimate, asymptotically we can determine $Z$ almost surely.
Hence,  we deduce deduce that $ H(Z|X^n,\Delta,D=d) = o(1)~ a.s.$
Therefore, if $H(\mb{\hat{w}}_d) \lesssim \frac{d}{2} \log n$, then
\begin{equation}
I(X^n;Z|\Delta,D=d) = H(\mb{\hat{w}}_d) + o(1).
\end{equation}
%\vspace{-.03in}
%\begin{lemma}
%If $2^{H(\mb{w})} = O\left(n^{\frac{\check{d}}{2}(1-\epsilon)}\right)$ for some $\epsilon>0$, then $H(Z|X^n,\Delta) = %o(1)$.
%\end{lemma}
%\noindent

To complete the proof of the theorem, we need to show that if $H(\mb{\hat{w}}_d) \gtrsim \frac{d}{2} \log n$, we have $I(X^n;Z|\Delta,D=d) = \bar{R}_{n,d} + o(1)$. In this case, $K \to \infty$ as $n \to \infty$, and hence, for any $\epsilon>0$, there exists a subset $\Delta'_d$ of the $K_d$ vectors of the $d$-dimensional parameter vectors indexed with $K'_d$, with normalized weight vector $\mb{\hat{u}}_d$, such that
\begin{equation}
(1-2\epsilon) \bar{R}_{n,d}  < H(\mb{\hat{u}_d}) < (1-\epsilon) \bar{R}_{n,d}. 
\end{equation}
Let $\mathbb{I}_{\Delta'_d}$ denote the indicator function of the subset $\Delta'_d$. It is straightforward to show that 
\begin{align}
I(X^n;Z|\Delta,D=d) &\geq I(X^n;Z|\mathbb{I}_{\Delta'_d}, \Delta,D=d)\\
&\geq I(X^n;Z|\mathbb{I}_{\Delta'_d}=1, \Delta,D=d).
\end{align}
%where $\mb{\hat{u}}_d$ denotes the weight vector of those $K'_d$ parameter vectors renormalized to sum to one.
Note that $\bar{R}_{n,d}\sim \frac{d}{2}\log n$ and hence $H(\mb{\hat{u}_d}) \lesssim \frac{d}{2}\log n$.
Therefore, we have
$I(X^n;Z|\mathbb{I}_{\Delta'_d}=1,\Delta,D=d) \geq (1-2\epsilon) \bar{R}_{n,d} + o(1)$ almost surely. On the other hand, we also have
\begin{equation}
I(X^n;Z|\Delta,D=d)\leq I(X^n;\theta^{(Z)}|D=d) = \bar{R}_{n,d}.
\end{equation}
Hence, we deduce that $I(X^n;Z|\Delta,D=d) = \bar{R}_{n,d}+ o(1)$ almost surely, completing the proof.
\end{IEEEproof}

\begin{IEEEproof}[Proof of Theorem~\ref{thm:Gallager2}]
The equivalence of the average minimax redundancy and the average maximin redundancy and the channel capacity above is a direct consequence of Theorem 5 of Gallager in~\cite{Gallager-source-coding}.
% Now, it only remains to show that the capacity achieving prior on the set $\Delta$ is such that for all $i \neq j$, $\theta^{(i)}$ and $\theta^{(j)}$ are independent and each $\theta^{(i)}$ follows Jeffreys' prior on the $d_i$-dimensional space $\Lambda_{d_i}$.
Next, let $\theta^{(i)}$ and $\theta^{(j)}$ be independently chosen according to Jeffreys' prior on the $d_i$-dimensional space $\Lambda_{d_i}$. Then, if $H(\hat{\mb{w}}_d) \lesssim \frac{d}{2} \log n$ almost surely, as $n \to \infty$, we have $\theta^{(i)}$ and $\theta^{(j)}$ are $\omega(\frac{1}{\sqrt{n}})$ apart. Hence, this choice will maximize the mutual information asymptotically almost surely. On the other hand, if $H(\mb{w}_d)$, almost surely you have too many parameter vectors that you cannot discriminate them, and hence, the mutual information is almost surely asymptotically vanishing regardless of how they are distributed. 
% It is straightforward to demonstrate that the average redundancy achieved using this prior is an upper limit on the average redundancy, and hence, it is the capacity achieving prior.
\end{IEEEproof}

\begin{IEEEproof}[Proof of Theorem~\ref{thm:UCOMP}]
%\subsection{}
\label{proof:thm:UCOMP}
In light of~\eqref{eq:channel:1}, we would need to derive $I(X^n; \Delta)$. Observe that by the chain rule we have
\begin{align}
I(X^n; \Delta,Z,D) &= I(X^n;\Delta)\nonumber\\
& + I(X^n;D|\Delta)\nonumber\\
&+ I(X^n;Z|D,\Delta)
\label{eq:expand1}
\end{align}
where $D$ is the random dimension of the source parameter vector. By applying the chain rule in a different order we get
\begin{align}
I(X^n; \Delta,Z,D) &= I(X^n;D)\nonumber\\
& + I(X^n;Z|D)\nonumber\\
&+ I(X^n;\Delta|Z,D)
\label{eq:expand2}
\end{align}
Note that $I(X^n;Z|D) = 0$ as the random vector $X^n$ would not decrease the uncertainty in the index of the source $Z$ as there is no information about the source parameter vectors. Next, consider $I(X^n;D).$ In light of~\cite{Csiszar06, optimal_Markov_order,kieffer_order} $D$ is the random dimension of the signal can be determined uniquely as $n$ grows to infinity, i.e., $\lim_{n \to \infty} H(D|X^n) = 0$. Hence, 
\begin{equation}
I(X^n;D) = H(D) - H(D|X^n) = H(D) + o(1)  = H(\mb{v}) + o(1).
\end{equation}
Similarly, $I(X^n;D|\Delta) =  H(\mb{v}) + o(1)$ as $H(D|\Delta) = H(D)$.
Further, $I(X^n;Z|D,\Delta)$ is calculated in the proof of Theorem~\ref{thm:entropy_rate}.
Finally, to derive $I(X^n;\Delta|Z,D)$ note that the parameter vectors are chosen independently, and hence, we have $I(X^n;\Delta|Z,D) = I(X^n;\theta^{(Z)}|Z,D)$. 
On the other hand, as each of the unknown parameter vectors follow Jeffreys' prior, we have $I(X^n;\theta^{(Z)}|Z=z,D=d) = \bar{R}_{n,d}$. 
Thus,
%\vspace{-.04in}
\begin{align}
I(X^n;\theta^{(Z)}|Z,D) &= \sum_{d=1}^{d_\text{max}} v_d \sum_{i=1}^\KK \hat{w}_{d,i} I(X^n;\theta^{(Z)}|Z=z) \\ 
& = \sum_{d=1}^{d_\text{max}} v_d  \bar{R}_{n,d}.
%\vspace{-.04in}
\end{align}
By combining~\eqref{eq:expand1} and~\eqref{eq:expand2} and the above, we arrive at the desired result.
\end{IEEEproof}

\begin{IEEEproof}[Proof of Theorem~\ref{thm:UCOM}]
%\vspace{-.03in}
%\subsection{}
\noindent In the case of {\UUCM}, we need to derive $I(X^n; \Delta | \mb{Y}^{n,T})$. Using the chain rule we have the following.
\begin{align}
I(X^n; \Delta, \mb{S}, Z, D| \mb{Y}^{n,T}) &= I(X^n; \Delta | \mb{Y}^{n,T})\nonumber\\
&+ I(X^n; \mb{S}, Z, D| \mb{Y}^{n,T}, \Delta).
\label{eq:expand3}
\end{align}
On the other hand, $(X^n,Z,D)$ is independent of $ (\mb{Y}^{n,T}, \mb{S})$  given $\Delta$. Hence, 
\begin{equation}
I(X^n; \mb{S}, Z, D| \mb{Y}^{n,T}, \Delta) = I(X^n; Z, D| \Delta),
\end{equation}
which has been characterized in the proof of Theorem~\ref{thm:UCOMP}. Applying the chain rule in a different order, we get
\begin{align}
I(X^n; \Delta, \mb{S}, Z, D| \mb{Y}^{n,T}) &= I(X^n;  \mb{S}, Z, D | \mb{Y}^{n,T})\nonumber\\
&+ I(X^n; \Delta | \mb{Y}^{n,T},  \mb{S}, Z, D).
\label{eq:expand4}
\end{align}
Now, considering $I(X^n;  \mb{S}, Z, D | \mb{Y}^{n,T})$ observe that
\begin{align}
I(X^n;  \mb{S}, Z, D | \mb{Y}^{n,T})&= I(X^n;  Z, D | \mb{Y}^{n,T}) \nonumber\\
&+ I(X^n;  \mb{S} | \mb{Y}^{n,T}, Z, D) .
\end{align}
Observe that $I(X^n;  Z, D | \mb{Y}^{n,T})$ can be made arbitrarily close to $I(X^n;  Z, D | \Delta)$ with $T$, i.e., $\forall \delta ~\exists T_0$ such that for $T>T_0$,
\begin{equation}
I(X^n;  Z, D | \mb{Y}^{n,T}) - I(X^n;  Z, D | \Delta) < \delta,
\end{equation}
and $I(X^n;  Z, D | \Delta)$ is characterized in the proof of Theorem~\ref{thm:UCOMP}.
Further note that
$I(X^n;  \mb{S} | \mb{Y}^{n,T}, Z, D)$ can be made arbitrarily small with $T$, i.e., $\forall \delta ~\exists T_1$ such that $I(X^n;  \mb{S} | \mb{Y}^{n,T}, Z, D) <\delta$ for $T>T_1$. 

Now, we only need to derive $I(X^n; \Delta | \mb{Y}^{n,T},  \mb{S}, Z, D)$ in~\eqref{eq:expand4}. We have
\begin{align}
&I(X^n; \Delta | \mb{Y}^{n,T},  \mb{S}, Z, D) = I(X^n; \theta^{(Z)} | \mb{Y}^{n,T},  \mb{S}, Z, D) \nonumber\\
&+ \sum_{\i \neq Z} I(X^n; \theta^{(i)} | \mb{Y}^{n,T}, \mb{S},Z,D, \theta^{(Z)}, \theta^{(1)},\ldots, \theta^{(i-1)})
\end{align}
All the summands of the second term are zero as $X^n$ is independent of all $\theta^{(i)}$ ($i \neq Z$) given $Z$. On the other hand, observe that 
\begin{align}
 I(X^n; \theta^{(Z)} | \mb{Y}^{n,T},  \mb{S}, Z, D) =  I(X^n; \theta^{(Z)} | \{Y^{n}(t)\}_{S(t) = Z}, Z, D).
 \label{eq:single-sourced}
\end{align}
The size of $\frac{1}{T}| \{Y^{n}(t)\}_{S(t) = Z}|$ can be made arbitrarily close to $\hat{w}_{D,Z}$ for sufficiently large $T$. On the other hand, in~\eqref{eq:single-sourced} the side information is from a single source. Hence, the mutual information can be obtained using Theorem~2 of~\cite{ISIT12_distributed}.
Combining all these pieces results in the desired result.
\end{IEEEproof}
\begin{IEEEproof}[Proof of Theorem~\ref{thm:UCPCM}]
Observe that 
\begin{align}
I(X^n; \Delta,\mb{S}, Z,D|\mb{Y}^{n,T}) &= I(X^n; \Delta|\mb{Y}^{n,T}) \nonumber\\
&+ I(X^n; \mb{S}, Z,D|\mb{Y}^{n,T}, \Delta).
\end{align}
The first term was characterized in Theorem~\ref{thm:UCOM} and the second term is equal to $I(X^n;Z,D|\Delta)$, which was derived in the proof of Theorem~\ref{thm:UCOMP}. Applying the chain rule in a different order we have
\begin{align}
I(X^n; \Delta,\mb{S}, Z,D|\mb{Y}^{n,T}) &= I(X^n;  \mb{S}, Z|\mb{Y}^{n,T}) \nonumber\\
&+ I(X^n; \Delta|\mb{Y}^{n,T},\mb{S}, Z)\nonumber\\
&+ I(X^n; D|\Delta, \mb{Y}^{n,T},\mb{S}, Z).
\end{align}
The second term in the expansion is what we are after while $I(X^n; D|\Delta, \mb{Y}^{n,T},\mb{S}, Z) =0$.
Considering $ I(X^n;  \mb{S}, Z|\mb{Y}^{n,T})$, we have
\begin{align}
I(X^n;  \mb{S}, Z|\mb{Y}^{n,T}) & = I (X^n;Z|\mb{Y}^{n,T})\nonumber\\
& + I(X^n;  \mb{S}|\mb{Y}^{n,T},Z).
\end{align}
Using similar arguments as in the proof of Theorem~\ref{thm:UCOM}, we can make $I (X^n;Z|\mb{Y}^{n,T})$ and $I(X^n;  \mb{S}|\mb{Y}^{n,T},Z)$  arbitrarily close to $I (X^n;Z|\Delta)$ and zero, respectively, for sufficiently large $T$. Putting all these facts together, we conclude that
\begin{equation}
I(X^n; \Delta|\mb{Y}^{n,T},\mb{S}, Z) = I(X^n; \Delta|\mb{Y}^{n,T}) + \delta,
\end{equation}
where $\delta$ can be made arbitrarily small for sufficiently large $T$, which completes the proof.
\end{IEEEproof}

\bibliographystyle{IEEEtran}
\bibliography{../../../phd_thesis_bib}

\end{document}